\documentclass[11pt,prd,letterpaper,showpacs,groupedaddress,superscriptaddress,nofootinbib,floatfix,preprintnumbers,tightenlines]{revtex4}
\usepackage{amssymb,amsmath,graphicx,xcolor,cancel}
\usepackage{bm,comment}

\newcommand{\lsim}{\raisebox{-0.7ex}{$\stackrel{\textstyle <}{\sim}$ }}

\begin{document}

%\begin{picture}(40,40)(20,-30)
%\includegraphics[scale=0.4]{mondrianlogo.pdf}
%\end{picture}

\vspace{-1cm}
\title{First lattice QCD study of the gluonic structure of light nuclei}

\author{Frank Winter}
\affiliation{Jefferson Laboratory, 12000 Jefferson Avenue, 
	Newport News, VA 23606, USA}

 \author{William Detmold} \affiliation{
Center for Theoretical Physics, 
Massachusetts Institute of Technology, 
Cambridge, MA 02139, USA}
 
 \author{Arjun S. Gambhir}
 \affiliation{Nuclear and Chemical Sciences Division, Lawrence Livermore National Laboratory, Livermore, CA, 94550}
  \affiliation{Nuclear Science Division, Lawrence Berkeley National Laboratory, Berkeley, CA, 94720}
 
\author{Kostas~Orginos}
\affiliation{Department of Physics, College of William and Mary, Williamsburg,
  VA 23187-8795, USA}
\affiliation{Jefferson Laboratory, 12000 Jefferson Avenue, 
Newport News, VA 23606, USA}

\author{Martin~J.~Savage}
\affiliation{Institute for Nuclear Theory, University of Washington, Seattle, WA 98195-1550, USA}

\author{Phiala~E.~Shanahan } 
\affiliation{Department of Physics, College of William and Mary, Williamsburg,
	VA 23187-8795, USA}
\affiliation{Jefferson Laboratory, 12000 Jefferson Avenue, 
	Newport News, VA 23606, USA}
\affiliation{
	Center for Theoretical Physics, 
	Massachusetts Institute of Technology, 
	Cambridge, MA 02139, USA}

\author{Michael~L.~Wagman} 
\affiliation{
	Center for Theoretical Physics, 
	Massachusetts Institute of Technology, 
	Cambridge, MA 02139, USA}
\affiliation{Department of Physics,
	University of Washington, Box 351560, Seattle, WA 98195, USA}
\affiliation{Institute for Nuclear Theory, University of Washington, Seattle, WA 98195-1550, USA}

\collaboration{NPLQCD Collaboration}

\date{\today}

\preprint{INT-PUB-17-035}
\preprint{MIT-CTP/4932}
\preprint{JLAB-THY-17-2540}

\pacs{11.15.Ha, % Lattice gauge theory
      12.38.Gc, % Lattice QCD calculations
}

%%%%%%%%%%%%%%%%%%%%%%%%%%%%%%%%%%%%%%%%%%%%%%
\begin{abstract}
The role of gluons in the structure of the nucleon and light nuclei is investigated using lattice quantum chromodynamics (QCD) calculations. 
The first moment of the unpolarised gluon distribution is studied in nuclei up to atomic number $A=3$ at quark masses corresponding to pion masses of $m_\pi\sim 450$ and $806$ MeV. 
Nuclear modification of this quantity defines a gluonic analogue of the EMC effect and is constrained to be less than $\sim 10$\% in these nuclei. This is consistent with expectations from phenomenological quark distributions and the momentum sum rule. 
In the deuteron, the combination of gluon distributions corresponding to the $b_1$ structure function is found to have a small first moment compared with the corresponding momentum fraction.
The first moment of the gluon transversity structure function is also investigated in the spin-1 deuteron, where a non-zero signal is observed at $m_\pi \sim 806$ MeV. This is the first indication of gluon contributions to nuclear structure that can not be associated with an individual nucleon. 
\end{abstract}

\maketitle

%%%%%%%%%%%%%%%%%%%%%%%%%%%%%%%%%%%%%%%%%%%%%%%
\section{Introduction}
\label{sec:intro}

The elucidation of the internal structure of hadrons and nuclei has been a defining focus of nuclear physics research for many decades.  The pursuit of this goal has led to the construction of successful models of nuclei, the discovery of the quark substructure of the proton, and eventually the discovery of QCD as the theory of interacting quarks and gluons that are the constituents of protons, neutrons, and nuclei. Experiments using electroweak probes have revealed detailed information about the quark structure of nucleons and nuclei. For example, the electromagnetic form factors are precisely determined, and the axial form factors are also constrained. The leading-twist quark distribution functions (PDFs) are known over significant kinematic ranges from deep-inelastic scattering (DIS) experiments, and generalised parton distributions (GPDs) and transverse momentum dependent quark distributions (TMDs) are under active study at JLab~\cite{Dudek:2012vr,Armstrong:2017wfw,Armstrong:2017zqr}, COMPASS~\cite{Abbon:2007pq} and RHIC~\cite{Aschenauer:2015eha}. In addition, there is conclusive evidence that the quark distributions in  nuclei are not simply determined by the sum of the quark distributions of the constituent nucleons, a feature known as the EMC effect \cite{Aubert:1983xm,Norton:2003cb,Geesaman:1995yd,Malace:2014uea,Arneodo:1992wf,Piller:1999wx,Hen:2013oha}. 
In contrast, knowledge of gluon distributions within hadrons and nuclei is sparse, as such quantities are only accessible through less precise means such as scaling violations in DIS and heavy vector meson production. While the unpolarised gluon distribution is reasonably well known for the proton, particularly at small Bjorken $x$, its uncertainty still limits the precision of theoretical predictions for Higgs production at the Large Hadron Collider~\cite{Baglio:2010ae} and little is known as to how this distribution is modified in nuclei. The gluon helicity is less constrained~\cite{Nakagawa:2017ljf,Franco:2016aet}, and the gluonic transversity \cite{Jaffe:1989xy}  (which exists for spin $J\ge1$ hadrons) has not yet been investigated experimentally. This latter quantity is also a leading-twist distribution and is uniquely gluonic as it does not mix with quark operators at leading twist \cite{Jaffe:1989xy,Hoodbhoy:1998vm}. Attendant to the quark GPDs and TMDs are a similar set of gluon GPDs~\cite{Diehl:2003ny,Berger:2001zb,Liuti:2014dda} and TMDs~\cite{Mulders:2000sh}. These quantities have not yet been investigated experimentally but are necessary components of a complete picture of the partonic structure of hadrons and nuclei.

One of the primary goals of the planned Electron-Ion Collider (EIC)~\cite{Geesaman:2015fha,Accardi:2012qut} is to measure gluonic aspects of the structure of nucleons and nuclei. Key quantities driving ongoing discussions of machine design are the gluon density and helicity distribution functions~\cite{Ent:2016lod,Boer:2016jnn,Pisano:2016jjf,Chudakov:2016otl} and new possibilities of probing `exotic gluons' \cite{Jaffe:1989xy} in nuclei (gluons not associated with a single nucleon) are being considered~\cite{JLAB.LOI}. 
The measurements that will eventually be obtained from an EIC will represent significant progress in understanding partonic structure~\cite{Cosyn:2016oiq}. Avenues that provide complimentary information in mapping out the QCD structure of hadrons and nuclei, and that inform the design and planning of the EIC, are important to pursue.
One such approach is lattice QCD (LQCD), which provides a non-perturbative method of calculating the hadronic physics that arises from QCD. 
As well as moments of collinear gluon distributions, LQCD studies can determine moments of gluon GPDs that are complementary to experimental scans of GPDs over limited kinematics. Recent progress in LQCD methods to study quark TMDs~\cite{Engelhardt:2015jba,Engelhardt:2015czw} and the $x$-dependence of quasi-distributions \cite{Ji:2013dva,Ma:2014jla,Radyushkin:2017cyf,Lin:2014zya,Alexandrou:2015rja,Orginos:2017kos} can be extended to the corresponding gluonic cases to provide important non-perturbative insight into these quantities. 

In this work, the gluonic structure of light nuclei is investigated for the first time. The first moments of the unpolarised gluon densities in spin-averaged nuclei (the gluon momentum fractions) and of the gluon transversity are studied in nuclear states of atomic number $A\le 3$ at two different sets of quark masses corresponding to pion masses of $m_\pi\sim450$ and 806 MeV. Clean signals are seen for the gluon momentum fraction at both sets of quark masses, while no signal is observed for the first moment of the difference between the unpolarised gluon PDF in the $j_z=\pm1$ and $j_z=0$ deuteron (the combination entering the $b_1(x)$ structure function~\cite{Hoodbhoy:1988am}). The gluonic transversity of the deuteron at $m_\pi\sim806$ MeV is resolved from zero at the level of three standard deviations. This latter result is the first indication of non-nucleonic `exotic gluon' contributions to nuclear structure. 
This study shows that these quantities can be accessed in LQCD, and includes estimates of the computational requirements to perform calculations that more directly impact phenomenology by resolving a gluonic analogue of the EMC effect and the first moment of the gluon transversity at lighter values of the quark masses.
These estimates also set the scale required for future investigations of more complicated aspects of gluonic structure including higher moments of PDFs, GPDS and TMDs.

\section{Gluonic structure of light nuclei}
\label{sec:lattice}

LQCD calculations of gluonic observables have been performed for only the simplest cases. 
There have been studies of the pion~\cite{Meyer:2007tm} and nucleon~\cite{Alexandrou:2013tfa,Horsley:2012pz} gluon momentum fractions and the gluon contributions to the nucleon mass and spin~\cite{Alexandrou:2017oeh,Horsley:2012pz,Yang:2016plb}, as well 
as the generalised form factors corresponding to the momentum fraction and first moment of the transversity distribution of the $\phi$ meson \cite{Detmold:2016gpy,Detmold:2017oqb}.
In this work, these studies are extended to light nuclei for the first time.

The dominant aspects of partonic structure of hadrons and nuclei are determined by the leading twist (twist-2) quark and gluon parton distributions 
(including collinear, transverse and generalised parton distributions). Moments of the gluon distributions are determined by matrix elements of local gluonic operators, namely
\begin{align}\label{eq:SIop}
\overline{\mathcal{O}}_{\mu_1\ldots\mu_n}(\mu) & =S\left[ G_{\mu_1\alpha}i\overleftrightarrow{D}_{\mu_3}\ldots i\overleftrightarrow{D}_{\mu_{n}}G_{\mu_2}^{\,\,\,\alpha}\right],
\\\label{eq:SDop}
\tilde{\mathcal{O}}_{\mu_1\ldots\mu_n}(\mu) & =S\left[ \tilde{G}_{\mu_1\alpha}i\overleftrightarrow{D}_{\mu_3}\ldots i\overleftrightarrow{D}_{\mu_n}G_{\mu_2}^{\,\,\,\alpha}\right],
\\\label{eq:TTop}
\mathcal{O}_{\nu_1\nu_2\mu_1\ldots\mu_n}(\mu) & =S\left[ G_{\nu_1\mu_1}i\overleftrightarrow{D}_{\mu_3}\ldots i\overleftrightarrow{D}_{\mu_n}G_{\nu_2\mu_2}\right],
\end{align}
corresponding to the unpolarised, helicity and transversity distributions, respectively. 
Here $G_{\mu\nu}$ is the field strength tensor and $\tilde{G}_{\mu\nu} =\frac{1}{2} \epsilon_{\mu\nu\alpha\beta}G^{\alpha\beta}$ is its dual, $\overleftrightarrow{D} = \frac{1}{2}\left(\overrightarrow{D}-\overleftarrow{D}\right)$, `$S$' denotes symmetrisation in the indices $\mu_i$ and subtraction of traces in all indices\footnote{Here, the normalisation convention of the symmetrisation and trace subtraction is $S[A_\mu B_\nu ] = \frac{1}{2}\left(A_\mu B_\nu + A_\nu B_\mu\right) - \frac{1}{4}g_{\mu\nu}A_\alpha B^\alpha$.}, and $\mu$ is the factorization and renormalisation scale. 
The operators with the fewest derivatives are expected to be the most well-determined in LQCD calculations and are the focus of this work.
For $n=2$, the unpolarised operator corresponds to the traceless part of the energy momentum tensor and its matrix elements encode the lightcone momentum fraction carried by gluons in the corresponding hadron or nucleus. The transversity operators require a double
helicity flip and their forward limit matrix elements vanish in targets of spin $J<1$~\cite{Jaffe:1989xy}, although they have non-zero off-forward matrix elements in targets of any spin~\cite{Hoodbhoy:1998vm,Belitsky:2000jk}.

The light nuclei investigated in this study are the deuteron ($d$), the dinucleon ($nn$), and $^3$He, or equivalently the triton ($^3$H), 
with spins $J=1,0,\frac{1}{2}$ (note that at the unphysically large quark masses used here, the dinucleon is a bound state~\cite{Orginos:2015aya,Beane:2012vq}). In this work, only the forward limit matrix elements of the lowest ($n$=2) spin-independent and transversity operators defined in the towers of Eqs.~\eqref{eq:SIop} and \eqref{eq:TTop} are computed (matrix elements of the $n=2$ gluonic helicity operator of Eq.~\eqref{eq:SDop} vanish by operator symmetries). The relevant decompositions of the forward nuclear matrix elements are as follows:

\noindent For spin-zero nuclei,
\begin{align}\label{eq:Jzero}
\langle h; p | \overline{\mathcal{O}}_{\mu_1\mu_2} | h; p \rangle &= b_2^{(h)}\!(\mu)\, S\left[p_{\mu_1} p_{\mu_2}\right]/ m_h,
\\
\langle h; p | {\mathcal{O}}_{\nu_1\nu_2\mu_1\mu_2} | h; p \rangle & = 0.
\end{align}
For spin-half nuclei,
\begin{align}\label{eq:Jhalf}
\langle h; p, s | \overline{\mathcal{O}}_{\mu_1\mu_2} | h; p, s \rangle &= b_2^{(h)}\!(\mu)\,  S\left[p_{\mu_1} p_{\mu_2}\right]/m_h,
\\
\langle h; p, s | {\mathcal{O}}_{\nu_1\nu_2\mu_1\mu_2} | h; p, s \rangle &= 0.
\end{align}
For spin-one nuclei \cite{Taneja:2011sy,Detmold:2016gpy,Jaffe:1989xy},
\begin{align}\label{eq:Jone}
\langle h; p, \epsilon | \overline{\mathcal{O}}_{\mu_1\mu_2} | h; p, \epsilon  \rangle &= b_2^{(h)}\!(\mu)\, S\left[p_{\mu_1} p_{\mu_2}\right]/m_h + c_2^{(h)}\!(\mu)\, S\left[m_h^2\epsilon _{\mu_1} \epsilon ^{*}_{\mu_2}-\frac{1}{3}p_{\mu_1}p_{\mu_2}\right]/m_h,
\\\label{eq:JoneDG}
\langle h; p, \epsilon  | {\mathcal{O}}_{\nu_1\nu_2\mu_1\mu_2} | h; p, \epsilon  \rangle & =a_2^{(h)}\!(\mu)\,S\left[ \left(p_{\nu_1} \epsilon_{\mu_1} - \epsilon_{\nu_1}p_{\mu_1}\right)(p_{\nu_2} \epsilon^{*}_{\mu_2}-\epsilon^{*}_{\nu_2} p_{\mu_2})\right]/m_h,
\end{align}
where the tensor structure multiplying $c_2^{(h)}$ transforms as $J=2$ under spatial rotations and vanishes when averaged over polarisation states.
In each case, $h$ labels the state, $m_h$ is its mass, $s$ denotes the spin of a $J=\frac{1}{2}$ particle, and $\epsilon^\mu$ represents the dimensionless polarisation vector of a $J=1$ particle\footnote{The polarisation vectors are defined by $
	\epsilon^\mu(\vec{p},\lambda) = \left( \frac{\vec{p}\cdot\vec{e}_\lambda}{m_h}, \vec{e}_\lambda + \frac{\vec{p}\cdot\vec{e}_\lambda}{m_h(m_h+E_h)}\vec{p}\right)$,
	where $\lambda=\{+,-,0\}$, $E_h=\sqrt{|\vec{p}|^2+m^2}$ is the energy of the state, and $
	\vec{e}_\pm = \mp \frac{1}{\sqrt{2}}(0,1,\pm i)$, $\vec{e}_0 = (1,0,0)$.}. As above, `$S$' denotes symmetrisation in the indices $\mu_i$ and and trace-subtraction in all indices.
With states normalised to $\langle h;p | h; p\rangle =\frac{E_h}{m_h}(2\pi)^3 \delta^3(\vec{0})$ (with additional delta functions for spin or polarisation for $J=\frac{1}{2}$ or $J=1$ states), the renormalisation-scale dependent constants $a_2^{(h)}$, $b_2^{(h)}$, $c_2^{(h)}$, which correspond to the $n=2$ Mellin moments of the relevant gluon distributions at scale $\mu$, are dimensionless.

\section{Lattice QCD details}
\label{sec:lqcd}

The present work is based on ensembles of gauge field configurations that have been used extensively in studies of the properties and interactions of light nuclei. The salient details for this study are summarised here; more complete discussions appear in Refs.~\cite{Orginos:2015aya,Beane:2012vq}. Two ensembles of gauge field configurations with different quark masses are used. In both cases the strange quark mass is close to its physical value and the light quark masses are chosen such that $m_\pi\sim 450$ MeV (Ensemble A) and $m_\pi\sim806$ MeV (Ensemble B). Note that the configurations used here are a subset of those used in previous works~\cite{Orginos:2015aya,Beane:2012vq}.
Each ensemble is generated using the L\"uscher-Weisz tadpole-improved gauge action~\cite{Luscher:1984xn} with one level of stout smearing~\cite{Morningstar:2003gk} and $N_f=2+1$ flavours of clover fermions~\cite{Sheikholeslami:1985ij}. The spatial and temporal lattice dimensions, $(L,T)$, are $(32, 96)$ for Ensemble A and $(32, 48)$ for Ensemble B. The lattice spacings and other details of the ensembles are shown in Table \ref{tab:params}.
\begin{table}
\begin{tabular}{c|c||cccccccc}
	\hline 
	& $L$ & $T$ & $a [fm]$ & $m_\pi$ [MeV] & $m_K$ [MeV] & $m_\pi L$ & $m_\pi T$ & $N_\text{cfg}$ & $N_\text{src}$ \\ 
	\hline 
	Ensemble A & 32 & 96 & 0.1167(16)  & 450(5) & 596(6) & 8.5 & 25.5 & 1550 & 42 \\ 
	Ensemble B & 32 & 48 & 0.1453(16) & 806(9) & 806(9)  &  19.0 & 28.5 & 508 & 416 \\ 
	\hline 
\end{tabular} 
\caption{Parameters of the ensembles of gauge-field configurations used in this work. $N_\text{src}$ measurements were performed on each of $N_\text{cfg}$  configurations for each ensemble. The lattice spacings were determined as described in~Refs.~\cite{Orginos:2015aya,Beane:2012vq}.
\label{tab:params}}
\end{table}

\subsection{Nuclei at $m_\pi=450$~MeV and $m_\pi=806$~MeV}
\label{sec:nuclei}

The two-point correlation functions
\begin{align}\label{eq:C2pt}
C^{\text{2pt}}_{h}(t,\vec{p}) = & \sum_{\vec x} e^{i\vec{p}\cdot \vec{x}}\langle \chi_h(t,\vec{x})\chi^\dagger_h(0,\vec{0})\rangle\,
\end{align}
were constructed for the nuclear states $h=N$, $nn$, $d$, and additionally for $^3$He for Ensemble B, 
for all possible spin states and with $\vec{p}=0$. Here $\chi_h$ is an appropriate interpolating operator for the state $h$. The interpolating operators used in the numerical work are those constructed in Ref.~\cite{Detmold:2012eu}, restricted to those containing only upper-spin components (in the Dirac spinor basis). 
In practice, correlation functions are built using baryon blocks~\cite{Basak:2005ir}, projected to well-defined momenta at the sink, as an intermediate step.
These blocks take the form
\begin{eqnarray}
{\cal B}_B^{ijk}({\vec p},t; x_0) & = & 
\sum_{\vec x} e^{i {\vec p}\cdot {\vec x}} 
S_i^{(f_1),i^\prime}({\vec x},t; x_0)
S_j^{(f_2),j^\prime}({\vec x},t; x_0)
S_k^{(f_3),k^\prime}({\vec x},t; x_0) 
b_{i^\prime j^\prime k^\prime}^{(B)},
\label{eq:NucBlock}
\end{eqnarray}
where $S^{(f)}$ is a quark propagator of flavour $f=\{u,d\}$, and the indices $i = \{1,\ldots, N_c N_s\}$ are combined spin-colour indices, where $N_c=3$ is the number of colours and $N_s=4$ is the number of spin components. 
The $f_i$ and the tensor $b^{(B)}$ depend on the spin and flavour of the baryon $B$. The quark propagators used in this construction originate from a source smeared smeared with 80 steps of gauge-invariant Gaussian smearing with parameter $\rho=3.5$ with stout-smeared gauge links, and have either a point (SP) sink, or a sink smeared using the same parameters as the source (SS).
For states with $A>1$, correlation functions projected to momenta $\vec{p}=(0,0,\pm 1)$ (in lattice units) were also constructed, and for Ensemble A, momenta $\vec{p}=(0,0,\pm2)$ were also used.

\begin{figure}
	\includegraphics[width=0.47\columnwidth]{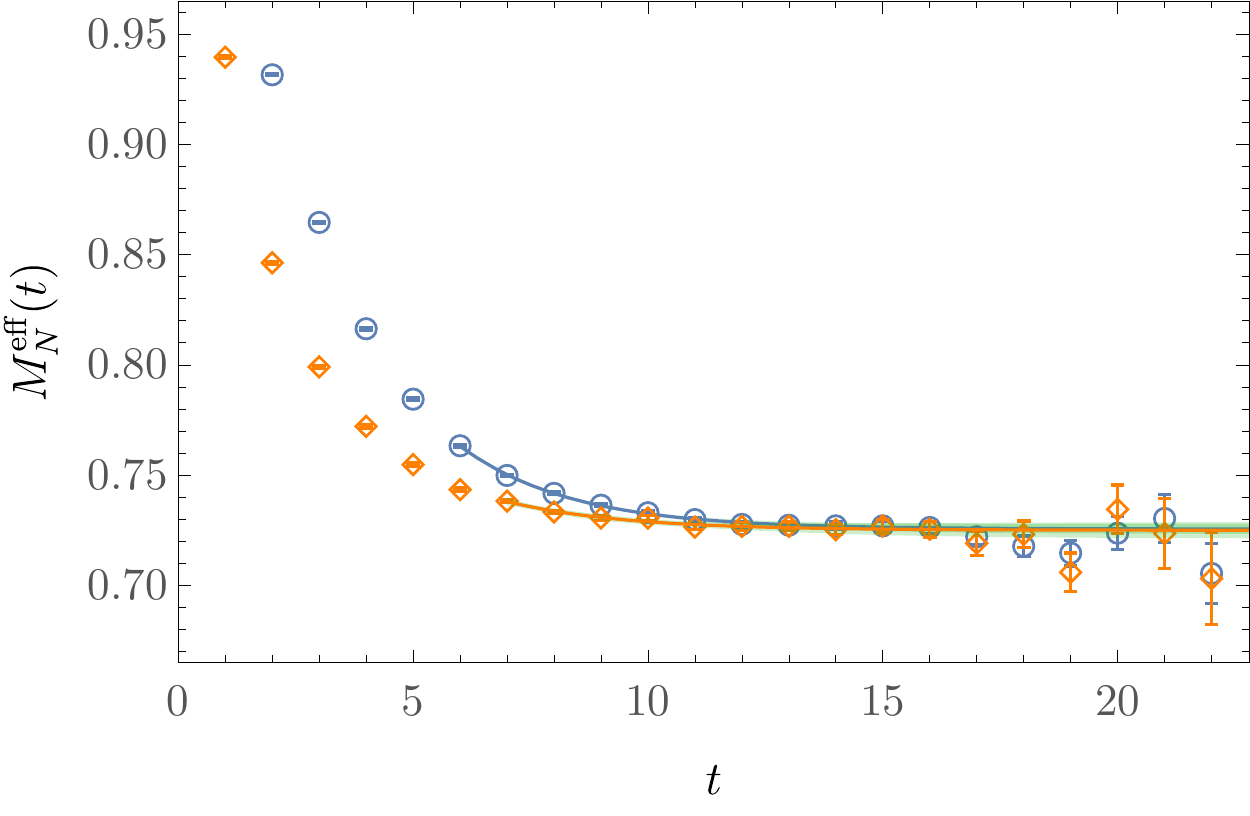}\qquad
	\includegraphics[width=0.47\columnwidth]{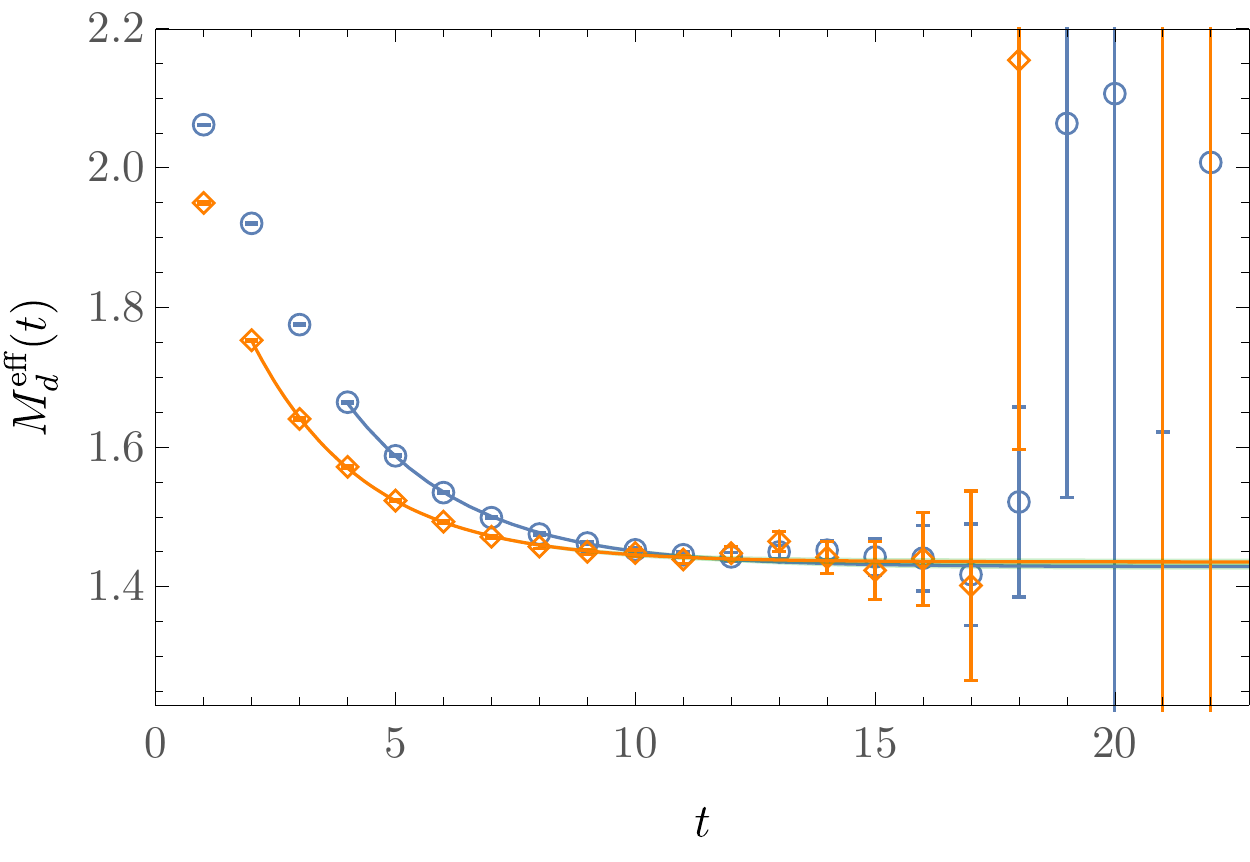} \\
	\includegraphics[width=0.47\columnwidth]{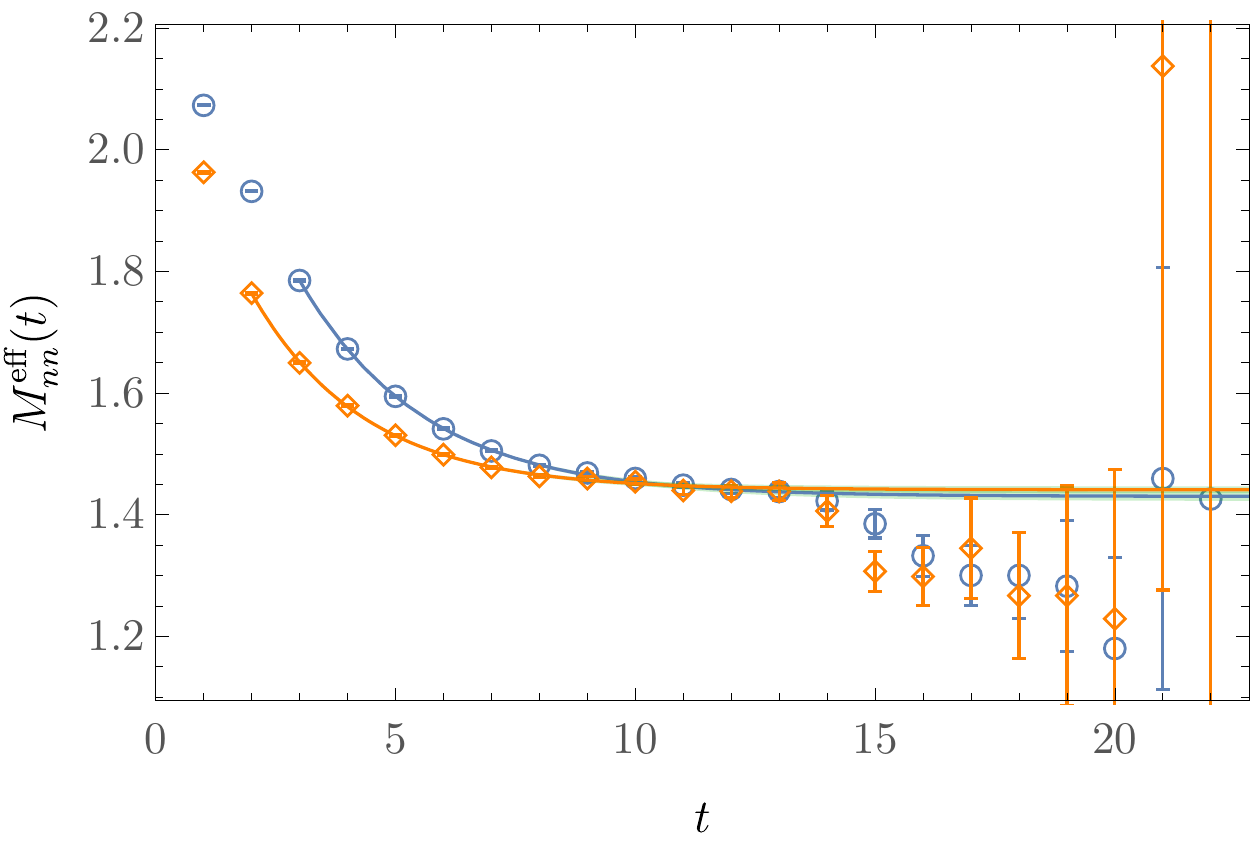}\qquad 
	\caption{Effective mass plots for the nuclear systems for Ensemble A (with $m_\pi \sim 450$ MeV). Blue circles and orange diamonds denote SP and SS source-sink smearing combinations, respectively. The bands show two-exponential fits to the correlation functions.}
	\label{fig:effmassA}
\end{figure}

\begin{figure}
	\includegraphics[width=0.47\columnwidth]{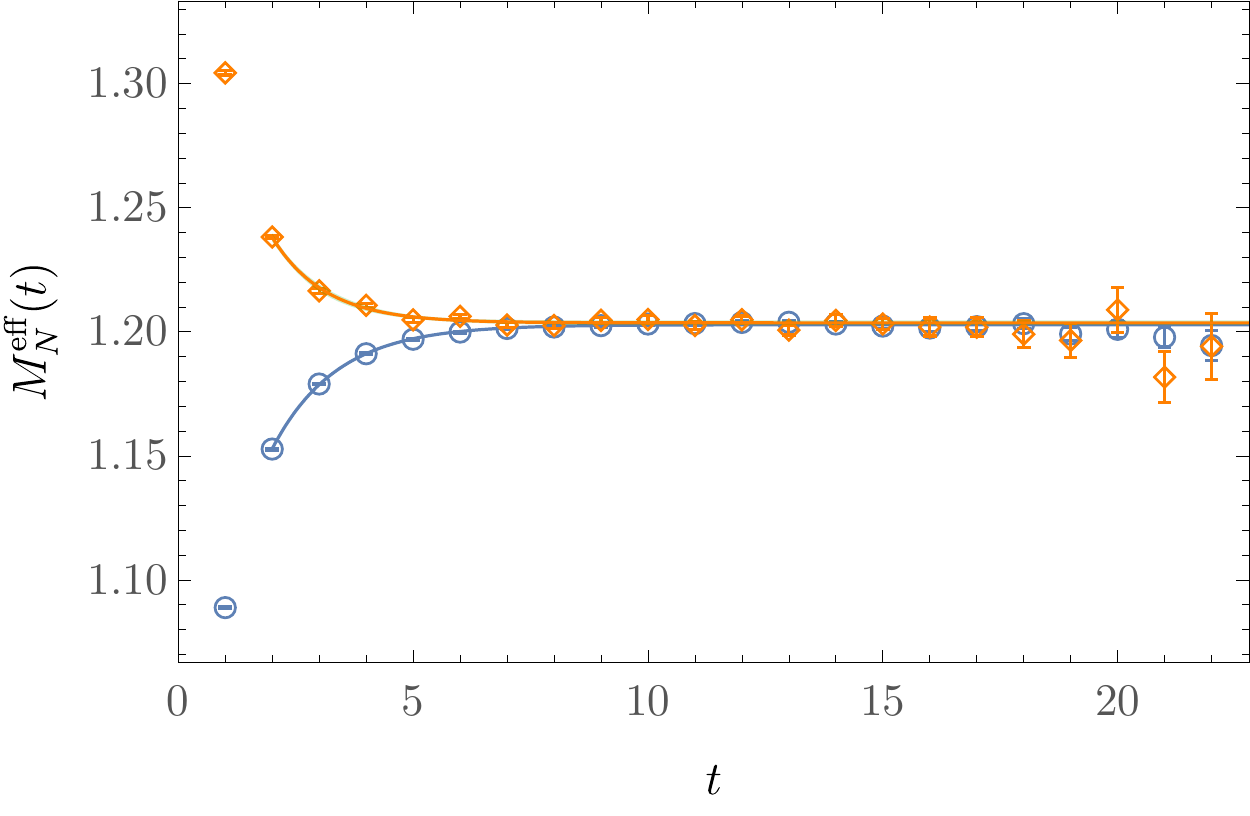}\qquad
	\includegraphics[width=0.47\columnwidth]{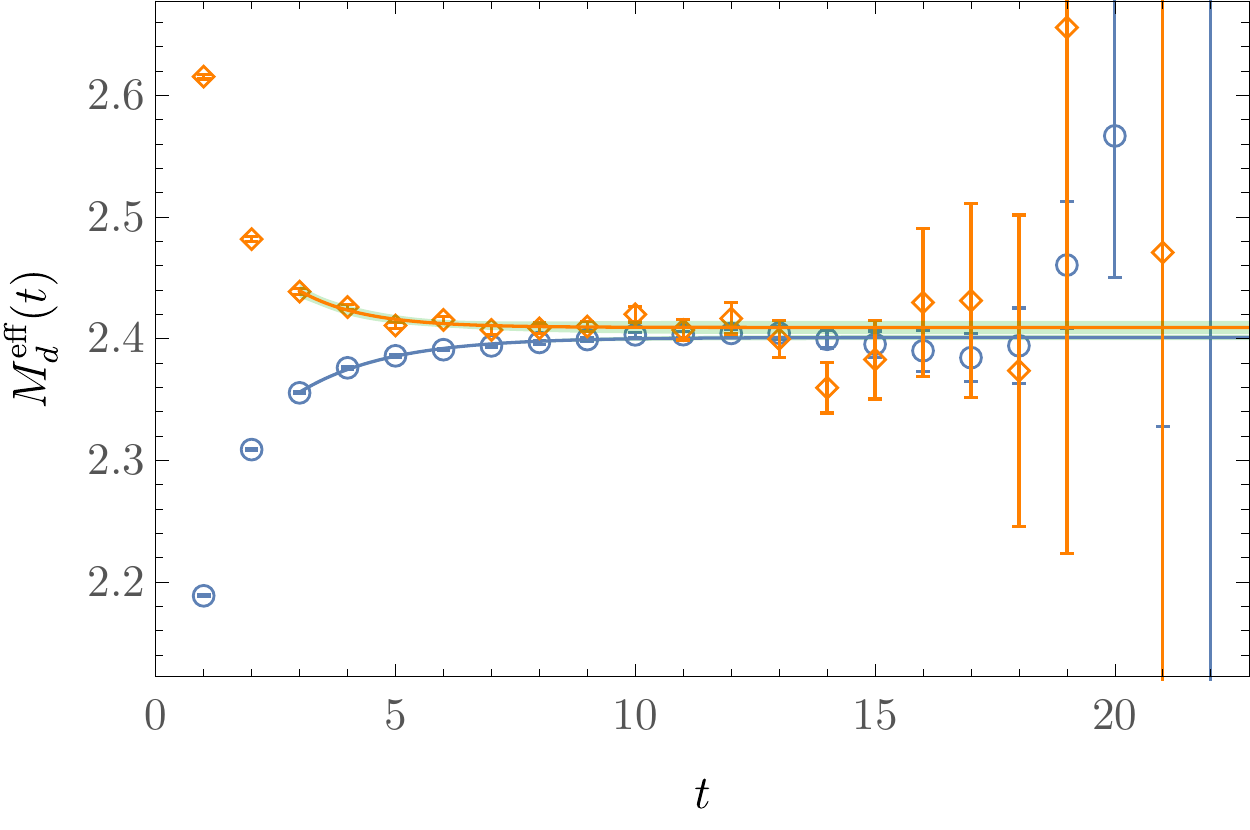} \\
	\includegraphics[width=0.47\columnwidth]{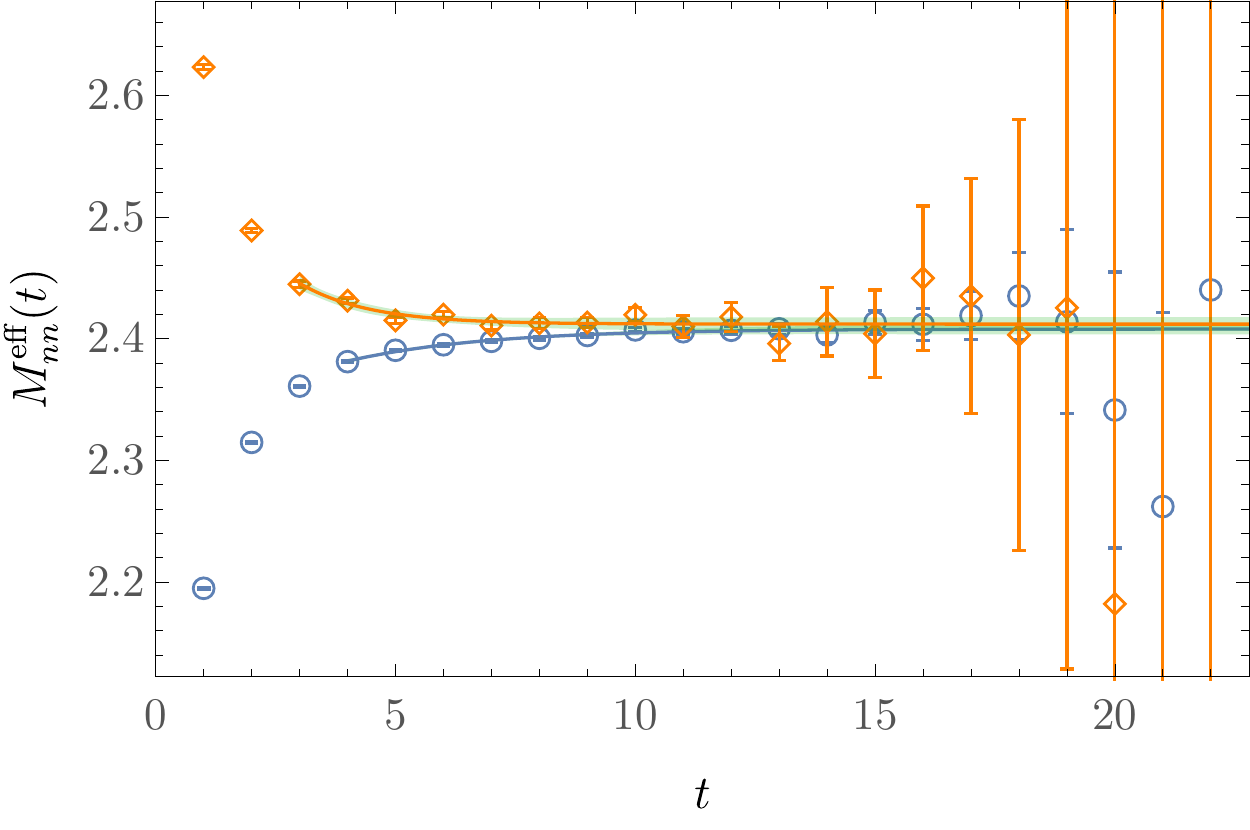}\qquad 
	\includegraphics[width=0.47\columnwidth]{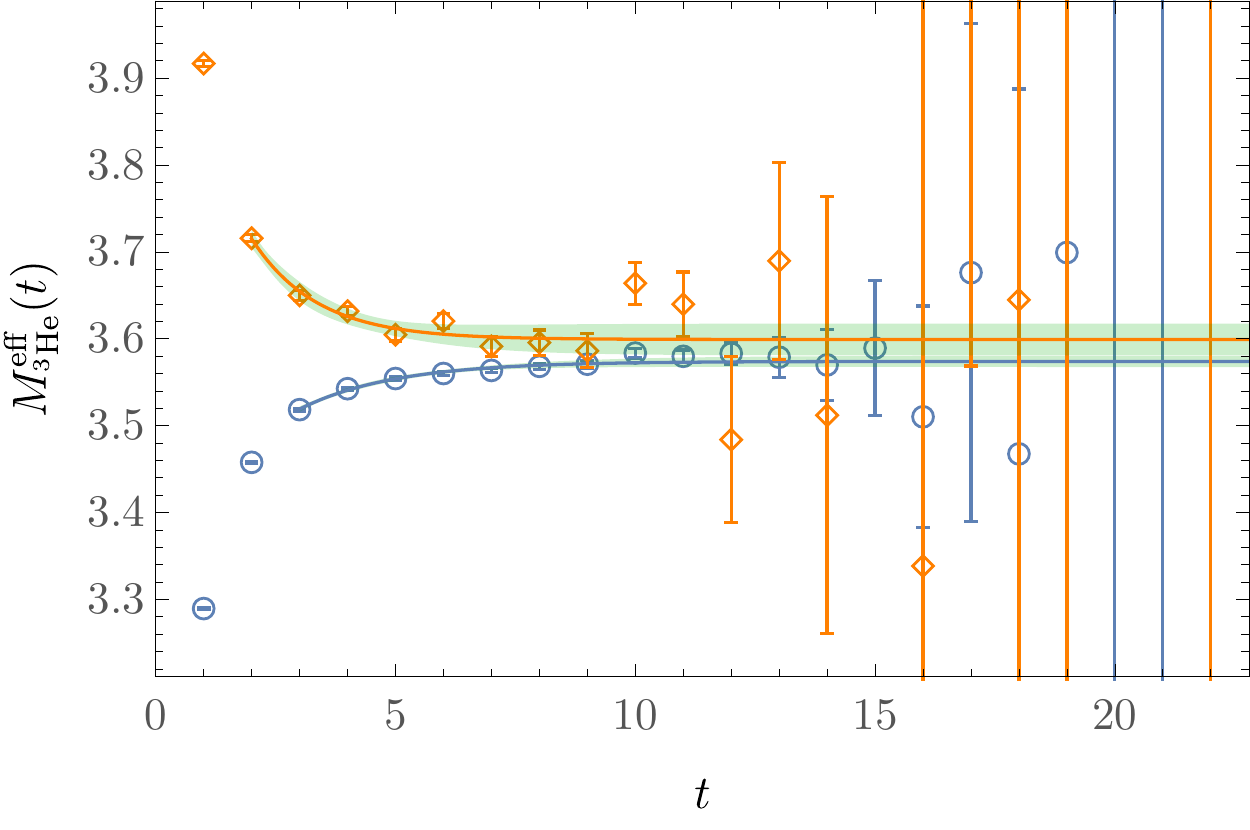}	
	\caption{Effective mass plots for the nuclear systems for Ensemble B (with $m_\pi \sim 806$ MeV). Blue circles and orange diamonds denote SP and SS  source-sink smearing combinations, respectively. The bands show two-exponential fits to the correlation functions.}
	\label{fig:effmassB}
\end{figure}

Effective mass ratios formed from the two point correlation functions for each state are shown in Fig.~\ref{fig:effmassA} for Ensemble A and in Fig.~\ref{fig:effmassB} for Ensemble B. The masses of all states determined using two-exponential fits to the correlation functions are in agreement with those found previously in higher-statistics calculations using the same ensembles in Refs.~\cite{Orginos:2015aya,Beane:2012vq}.
Moreover, the energy gaps between ground and first excited states extracted from the fits to data with SS and SP source-sink smearing combinations are entirely consistent for each nuclear state.
The downward fluctuation of the correlation function at timeslices $t\ge 15$ for the $nn$ state at 450~MeV that is evident in Fig.~\ref{fig:effmassB} was less pronounced in earlier higher-statistics calculations on the same ensemble~\cite{Orginos:2015aya}.
 Refs.~\cite{Wagman:2017tmp,Beane:2017edf} detail an extensive set of checks performed to verify that each extracted energy level corresponds to a bound ground state.

\subsection{Operators}

Three-point correlation functions are constructed by correlating the two-point functions discussed in the previous section with the relevant gluonic operators. 
The construction of these operators, and operator mixing considerations, are detailed below. 

\subsubsection{Spin-independent gluon operators}
\label{sec:SIops}

The operator related to the spin-independent gluon distributions, Eq.~\eqref{eq:SIop}, can be represented in discrete Euclidean space as 
\begin{eqnarray}
\overline{\mathcal{O}}^{(E)}_{\mu\nu} = G_{\mu\alpha}^{(E)} G_{\nu\alpha}^{(E)},
\end{eqnarray}
where the Euclidean-space field-strength tensor is related component-wise to the Minkowski definition by
\begin{align}
G^{(E)}_{ij} & = G_{ij} \text{   if } i,j\in \{1,2,3\},\\
G^{(E)}_{4j} & = (-i)G_{0j}.
\end{align}
Components of this operator are computed numerically using the clover field-strength tensor
\begin{equation}
G_{\mu\nu}^{(E)}(x) = \frac{1}{8}\left(P_{\mu\nu}(x)-P^\dagger_{\mu\nu}(x)\right),
\end{equation}
derived from the combination of plaquettes
\begin{align}\nonumber
P_{\mu\nu}(x) = & \,U_\mu(x) U_\nu(x+\mu)U_\mu^\dagger(x+\nu) U_\nu^\dagger(x)\\\nonumber
& + U_\nu(x)U^\dagger_\mu(x-\mu+\nu)U^\dagger_\nu(x-\mu)U_\mu(x-\mu)\\\nonumber
& + U^\dagger_\mu(x-\mu)U^\dagger_\nu(x-\mu-\nu)U_\mu(x-\mu-\nu)U_\nu(x-\nu)\\
& + U^\dagger_\nu(x-\nu)U_\mu(x-\nu)U_\nu(x-\nu+\mu)U^\dagger_\mu(x),
\end{align}
with $U_\mu(x)$ being the gauge field link in direction $\mu$, originating at site $x$.
Calculations of the gluon operators are performed with gradient flow~\cite{Luscher:2010iy} applied to the links in order to reduce statistical fluctuations. The results shown in later sections use operators flowed to a total time of 1 in lattice units using an integration step size of 0.01. The effects of using different levels of HYP smearing~\cite{Hasenfratz:2001hp}, as well as gradient flow, were investigated in a similar calculation of the gluonic structure of the $\phi$ meson~\cite{Detmold:2016gpy}, with consistent results found for the different smearing implementations.

In the lattice geometry, operators are classified according to their transformations under the hypercubic group, H(4). Basis operators from two irreducible H(4) representations that subduce from the spin-independent gluon operator, Eq.~\eqref{eq:SIop}, are considered here. These do not suffer from mixing with same or lower-dimensional operators induced by the breaking of the rotational symmetry. The construction of these operators is detailed in Refs.~\cite{Detmold:2016gpy,Gockeler:1996mu}, and the same notation as in those works is used here. 
For the first representation considered, named $\tau_1^{(3)}$, the basis vectors are chosen as:
\begin{align}
\overline{\mathcal{O}}_{{1,1}}^{(E)} = & \frac{1}{2}\left(\overline{\mathcal{O}}^{(E)}_{11}+\overline{\mathcal{O}}^{(E)}_{22}-\overline{\mathcal{O}}^{(E)}_{33}-\overline{\mathcal{O}}^{(E)}_{44}\right), \\
\overline{\mathcal{O}}_{{1,2}}^{(E)} = & \frac{1}{\sqrt{2}} \left(\overline{\mathcal{O}}^{(E)}_{33}-\overline{\mathcal{O}}^{(E)}_{44}\right),\\
\overline{\mathcal{O}}_{{1,3}}^{(E)} = & \frac{1}{\sqrt{2}} \left(\overline{\mathcal{O}}^{(E)}_{11}-\overline{\mathcal{O}}^{(E)}_{22}\right).
\end{align}
For $\tau_3^{(6)}$ the basis vectors that are used are:
\begin{align}
\overline{\mathcal{O}}_{{2,\mu\nu}}^{(E)} = & \frac{1}{\sqrt{2}} \left(\overline{\mathcal{O}}^{(E)}_{\mu\nu}+\overline{\mathcal{O}}^{(E)}_{\nu\mu}\right), \hspace{2mm} 1\le \mu < \nu \le 4.
\end{align}
While matrix elements constructed using operators in different representations may yield reduced matrix elements (i.e., $a^{(h)}_2$, $b_2^{(h)}$, $c_2^{(h)}$ in Eqs.~\eqref{eq:Jzero}--\eqref{eq:Jone}) that differ by lattice-spacing artefacts, all should give the same results in the continuum limit. Studying multiple irreducible representations of H(4) containing operators that have the same continuum limit gives some insight into the significance of such discretisation artefacts, while using multiple vectors from each irreducible basis allows averaging over correlation functions to reduce statistical fluctuations in the results.

\subsubsection{Gluon transversity operators}

The lowest moment of the gluon transversity distribution in a given hadron or nucleus is determined by the corresponding matrix element of the Euclidean-space operator
\begin{eqnarray}
\mathcal{O}^{(E)}_{\mu\nu\mu_1\mu_2} = G_{\mu\mu_1}^{(E)} G_{\nu\mu_2}^{(E)}.
\end{eqnarray}
In this work, operators in a single irreducible representation are considered. Operators from other irreducible representations that do not mix with same or lower dimension operators have forward matrix elements that are statistically less-well determined, or vanish, for the boosts considered.

\noindent
The basis vectors in the $\tau_1^{(2)}$ representation are\footnote{Note that for the particular boosts considered here, forward matrix elements of $\mathcal{O}_{1,2}^{(E)}$ vanish.}~\cite{Gockeler:1996mu}:
\begin{align}\nonumber
\mathcal{O}_{{1,1}}^{(E)} = & \frac{1}{8\sqrt{3}}\left(-2 {\mathcal{O}}^{(E)}_{1122}+{\mathcal{O}}^{(E)}_{1133}+{\mathcal{O}}^{(E)}_{1144}+{\mathcal{O}}^{(E)}_{2233}+{\mathcal{O}}^{(E)}_{2244}-2 {\mathcal{O}}^{(E)}_{3344} \right), \\
\mathcal{O}_{{1,2}}^{(E)} = & \frac{1}{8} \left({\mathcal{O}}^{(E)}_{1144}+{\mathcal{O}}^{(E)}_{2233}-{\mathcal{O}}^{(E)}_{1133}-{\mathcal{O}}^{(E)}_{2244}\right).
\end{align}
These operators are constructed in the same way as detailed above for the spin-independent case.

\subsubsection{Mixing and Renormalisation}
\label{sec:renorm}

Matrix elements calculated in lattice QCD must be renormalised in order to correspond to well-defined quantities in the continuum limit. 
To make contact with phenomenological PDF parameterisations fit to experimental data, the $\overline{\text{MS}}$ renormalisation scheme is preferred. Ref.~\cite{Alexandrou:2016ekb} showed that this renormalisation is less than a 10\% effect for the spin-independent operator, matching to $\overline{\text{MS}}$ at $\mu^2=4$~GeV$^2$. 
The renormalisation has not been computed for the particular action used in the present calculation, but will cancel, up to small effects of mixing, in ratios of matrix elements of the gluon operators in light nuclei to the nucleon. 

Precisely, the spin-independent gluonic operator with $n = 2$, Eq.~\eqref{eq:SIop}, mixes with flavour-singlet quark operator $\overline{\mathcal{Q}}^{(E)}_{\mu_1\mu_2}=\sum\limits_{f=\{u,d,s\}}S\left[\overline{\psi}_f\gamma_{\mu_1} \overleftrightarrow{D}_{\mu_2}\psi_f\right]$, where $\psi_f$ is a quark field of flavour $f$. This mixing takes the form $\overline{\mathcal{O}}^{(E)} = Z^{gg} \mathcal{\overline{O}}^{\text{ren.} }+ Z^{gq}\mathcal{\overline{Q}}^{\text{ren.}}$, where lattice operators are denoted by the superscript $(E)$, while renormalised Euclidean-space operators (in any particular scheme) carry the superscript `ren'. It was shown in Ref.~\cite{Alexandrou:2016ekb}, using a one-loop perturbative renormalisation procedure and a similar action to the one used here, that this mixing is a few-percent effect. 
Contamination in the ratio
\begin{align}\nonumber
\frac{\langle h | \overline{\mathcal{O}}^{(E)}|h\rangle}{\langle N | \overline{\mathcal{O}}^{(E)}|N\rangle} & =  \frac{Z^{gg}\langle h | \overline{\mathcal{O}}^{\text{ren.}} | h \rangle + Z^{qg}\langle h | \overline{\mathcal{Q}}^{\text{ren.}} | h \rangle }{Z^{gg}\langle N | \overline{\mathcal{O}}^{\text{ren.}} | N \rangle + Z^{qg}\langle N | \overline{\mathcal{Q}}^{\text{ren.}} | N \rangle } \\
& \approx \frac{\langle h | \overline{\mathcal{O}}^{\text{ren.}}|h\rangle}{\langle N | \overline{\mathcal{O}}^{\text{ren.}}|N\rangle}\left(1 + \frac{Z^{qg}}{Z^{gg}}\left[ \frac{\langle h | \overline{\mathcal{Q}}^{\text{ren.}} | h \rangle}{\langle h | \overline{\mathcal{O}}^{\text{ren.}} | h \rangle} - \frac{\langle N | \overline{\mathcal{Q}}^{\text{ren.}} | N \rangle}{\langle N | \overline{\mathcal{O}}^{\text{ren.}} | N \rangle}\right]\right)
\end{align}
is then suppressed both by the mixing, i.e., $Z^{qg}/Z^{gg}\sim \mathcal{O}(\alpha_s) $, and by the small difference between nuclear and nucleon matrix elements (the term in square brackets).
These ratios are key quantities of interest in the present study, as deviations of their values from unity would reveal a gluonic analogue of the EMC effect.

In the transversity case, the relevant operator for $n$ = 2, Eq.~\eqref{eq:TTop}, does not mix with quark-bilinear operators of the same twist under renormalisation. This makes it a clean probe of gluonic structure, and the corresponding structure function is  a key target of upcoming experimental investigations of the gluon structure of nuclei~\cite{JLAB.LOI}.

\subsection{Matrix elements}
\label{sec:MEs}

Three-point correlation functions are constructed by taking the correlated product, configuration-by-configuration and source-location--by--source-location, of the two-point functions with the gluonic operators constructed as described in the previous sections. For a given operator $\mathcal{O}$, these can be expressed as:
\begin{align}\label{eq:C3pt}
C^\text{3pt}_{h}(t,\tau,\vec{p},\mathcal{O})  =  & \sum_{\vec x}\sum_{\vec y} e^{i\vec{p}\cdot \vec{x}} \left(\langle \chi_h(t,\vec{p}) \ \mathcal{O}(\tau,\vec{y})\ \chi_h^\dagger(0,\vec 0)\rangle - \langle \chi_h(t,\vec{p}) \ \chi_h^\dagger(0,\vec 0)\rangle \langle \mathcal{O}(\tau,\vec{y})\rangle \right), \\\nonumber
= & \, Z_h e^{-E_ht} \langle \vec{p},h|\mathcal{O}|\vec{p},h\rangle + Z_{h'} e^{-(E_h+\Delta_h)t} \langle  \vec{p},h'|\mathcal{O}|\vec{p},h'\rangle\\\nonumber
&+Z'_{h}e^{-(E_h+\Delta_h)\tau} e^{-E_h(t-\tau)}\langle \vec{p},h|\mathcal{O}|\vec{p},h'\rangle\\
&+Z'_{h'}e^{-(E_h+\Delta_h)(t-\tau)} e^{-E_h \tau}\langle \vec{p},h'|\mathcal{O}|\vec{p},h\rangle+\ldots,
\end{align}
where $\vec{p}$ labels the momentum of the state and the second term in Eq.~\eqref{eq:C3pt} corresponds to the subtraction of disconnected contributions\footnote{While this subtraction vanishes for the case of the gluon transversity considered in Section~\ref{sec:gt}, the correlated subtraction of zero is nevertheless found to cancel some of the noise in the numerical calculations.}. The energy of state $h$ is denoted $E_h$, and the energy gap to the first excited state is $\Delta_h$. Here $\chi_h$ is an appropriate interpolating operator for the state $h$, which additionally indexes the various polarisation and spin components of spin-1 or spin-$\frac{1}{2}$ states.
The second line follows by inserting complete sets of states to make the time-dependence of the function explicit, and holds for
$0\ll\tau\ll t\ll T$ (where $T$ denotes the time extent of the lattice). For the case $0\ll t \ll \tau\ll T$, $t$ is replaced by $(T-t)$ in the final line of the above expression and there is an additional multiplicative factor of $(-1)^{n_4}$, where $n_4$ is the number of temporal indices in the operator. Contributions from the first excited state, denoted by $h'$, are made explicit, while  the ellipsis corresponds to those from higher excitations (which are further exponentially suppressed), which are omitted.

The time-dependence of the two-point function in Eq.~\eqref{eq:C2pt}, for $t\ll T/2$, is
\begin{equation}
C^\text{2pt}_{h}(t,\vec{p})  = Z_h e^{-E_ht}+Z_{h'}e^{-(E_h+\Delta_h)t}+\ldots,
\end{equation}
where, again, exponentially-suppressed contributions from higher excited states are omitted. Forming a ratio of the three and two-point functions to cancel the leading exponential time dependence and $Z_h$ factors gives a quantity that reduces to the matrix element of interest plus exponentially-suppressed contamination from excited states:
\begin{align}\label{eq:rat}
R_{h}(t,\tau,\vec{p},\mathcal{O})&=\frac{C^\text{3pt}_{h}(t,\tau,\vec{p},\mathcal{O})}{C^\text{2pt}_{h}(t,\vec{p})}\pm\frac{ C^\text{3pt}_{h}(T-t,T-\tau,\vec{p},\mathcal{O})}{C^\text{2pt}_{h}(T-t,\vec{p})}\\\nonumber
&=\langle \vec{p},h|\mathcal{O}|\vec{p},h\rangle + \frac{Z_{h'}}{Z_h}\left(e^{-\Delta_h t}\left[\langle \vec{p},h'|\mathcal{O}|\vec{p},h'\rangle - \langle \vec{p},h|\mathcal{O}|\vec{p},h\rangle\right] \right) \\\label{eq:multiexpintermediate}
& \hspace{1.1cm}+\frac{Z'_{h'}}{Z_h}\left(e^{-\Delta_h (t-\tau)}\langle \vec{p},h'|\mathcal{O}|\vec{p},h\rangle + e^{-\Delta_h \tau}\langle \vec{p},h|\mathcal{O}|\vec{p},h'\rangle \right)+\ldots \\\label{eq:multiexp}
& = \langle \vec{p},h|\mathcal{O}|\vec{p},h\rangle + A e^{-\Delta_h t} + B e^{-\Delta_h(t-\tau)} + C e^{-\Delta_h \tau}+\ldots
\end{align}
for $\{t,\tau\} \ll T/2$, where the sign in the numerator on the right-hand side is $+ (-)$ for an even (odd) number of temporal indices in the operator $\mathcal{O}$. 
In arriving at Eq.~\eqref{eq:multiexpintermediate}, the contributions of excited states have been assumed to be small, enabling expansion of the denominators in Eq.~\eqref{eq:rat}.

\section{Gluon momentum fractions in light nuclei}
\label{sec:x}

The fraction of longitudinal momentum carried by gluons inside a given hadronic state is determined by matrix elements of the $n=2$ case of the spin-independent gluon operator $\overline{\mathcal{O}}$ defined in Eq.~(\ref{eq:SIop}). The reduced matrix elements $b_2^{(h)}$, defined in Eqs.~\eqref{eq:Jzero}, \eqref{eq:Jhalf}, \eqref{eq:Jone}, correspond precisely to this momentum fraction, i.e., the $x$-weighted integral of the unpolarised gluon distribution in the unpolarised hadron.
In the spin-1 deuteron, the reduced matrix element $c_2^{(d)}$ (see Eq.~\eqref{eq:Jzero}) corresponds to the first moment of the difference between the unpolarised gluon PDF in the $j_z=\pm 1$ and $j_z=0$ deuteron states~\cite{Hoodbhoy:1988am}.

The gluon momentum fractions in the nucleon and in light nuclei are determined from ratios $R_h$ of two and three-point correlation functions, constructed as described in Sec.~\ref{sec:MEs}. 
These ratios are first averaged over contributions from the different operators in the $\tau_1^{(3)}$ basis (see Sec.~\ref{sec:SIops}), over different boost momenta, and over the different spin and polarisation states for the $J=\frac{1}{2}$ and $J=1$ hadrons. To construct a quantity that determines the momentum fractions $b_2^{(h)}$, this averaging is performed including appropriate factors of momentum and mass through Eqs.~\eqref{eq:Jzero}, \eqref{eq:Jhalf} and \eqref{eq:Jone}, and any contributions with statistical uncertainties greater than 1.5 times the best-determined contribution are omitted from the average. 

\begin{figure}
	\includegraphics[height=0.67\textheight]{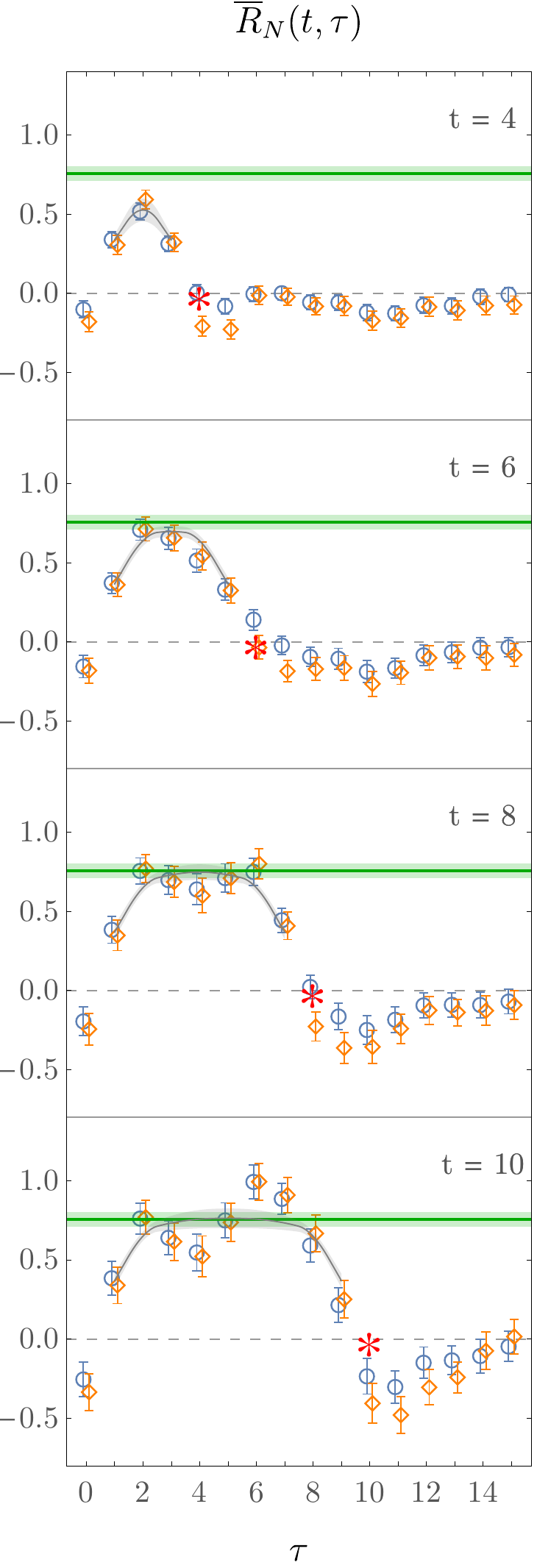}
	\includegraphics[height=0.67\textheight]{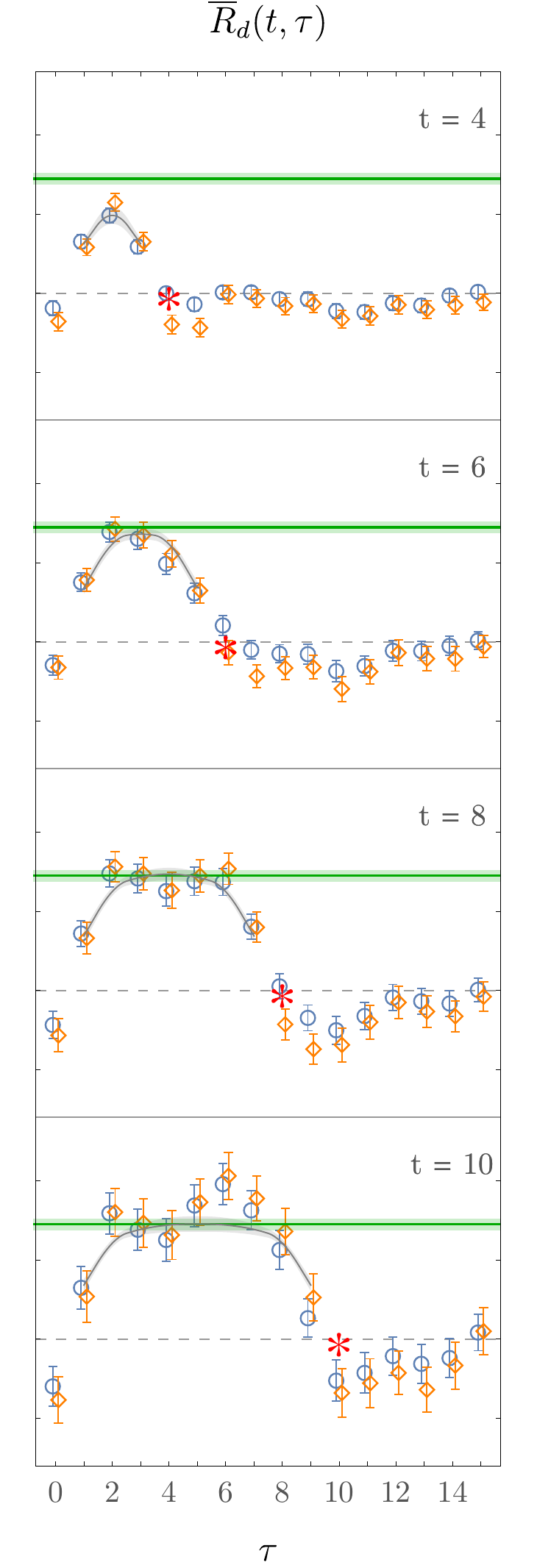}
	\includegraphics[height=0.67\textheight]{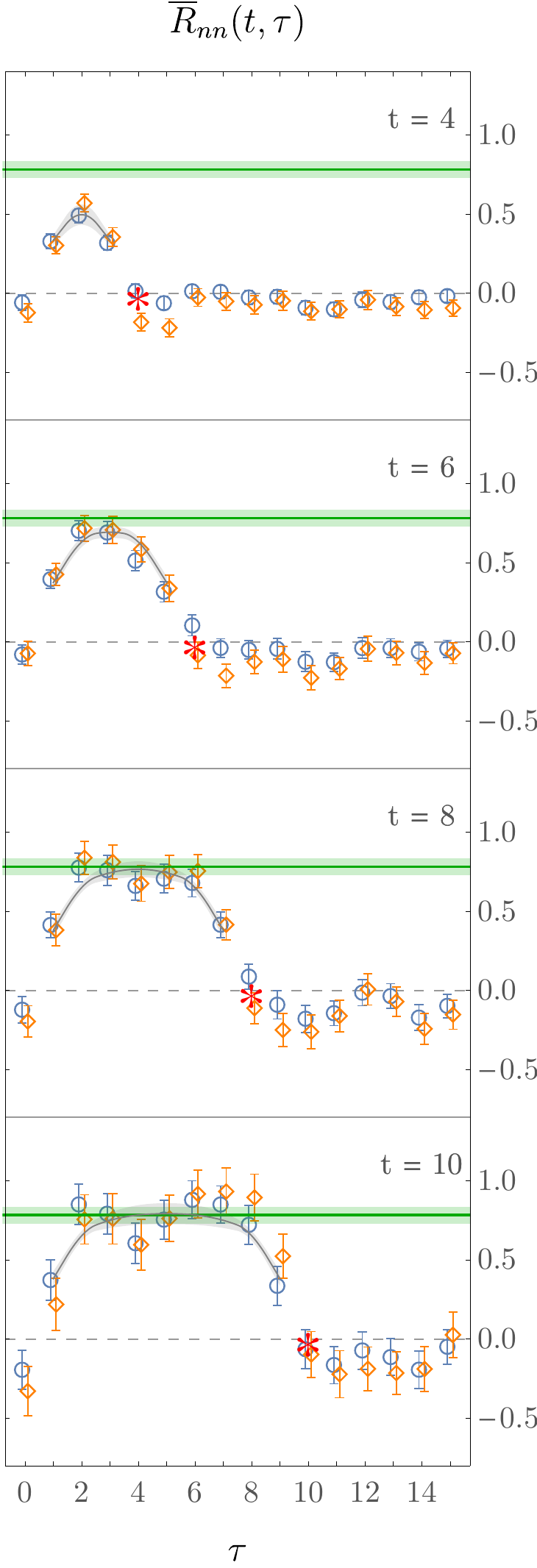}
	\caption{\label{fig:bumps450}Fits to the ratios $\overline{R}_h$ for $h$ = \{$N$, $d$, $nn$\}, averaged over operators in the $\tau_1^{(3)}$ irreducible representation and for different momenta with appropriate factors through Eqs.~\eqref{eq:Jhalf} and \eqref{eq:Jone} to extract $b_2^{(h)}$, for Ensemble A (with $m_\pi \sim 450$ MeV). Blue circles and orange diamonds denote SP and SS source-sink smearing combinations, respectively, while the red stars denote the sink times. The green bands (repeated on each panel) show the extracted results for the matrix elements from fits in the $t$--$\tau$ plane to the SP ratios using the exponential form of Eq.~\eqref{eq:rat}. The grey curves show those exponential fits. The scales on each figure are the same, and the light grey lines indicate the zero of each panel.}
\end{figure}

\begin{figure}
	\includegraphics[height=0.68\textheight]{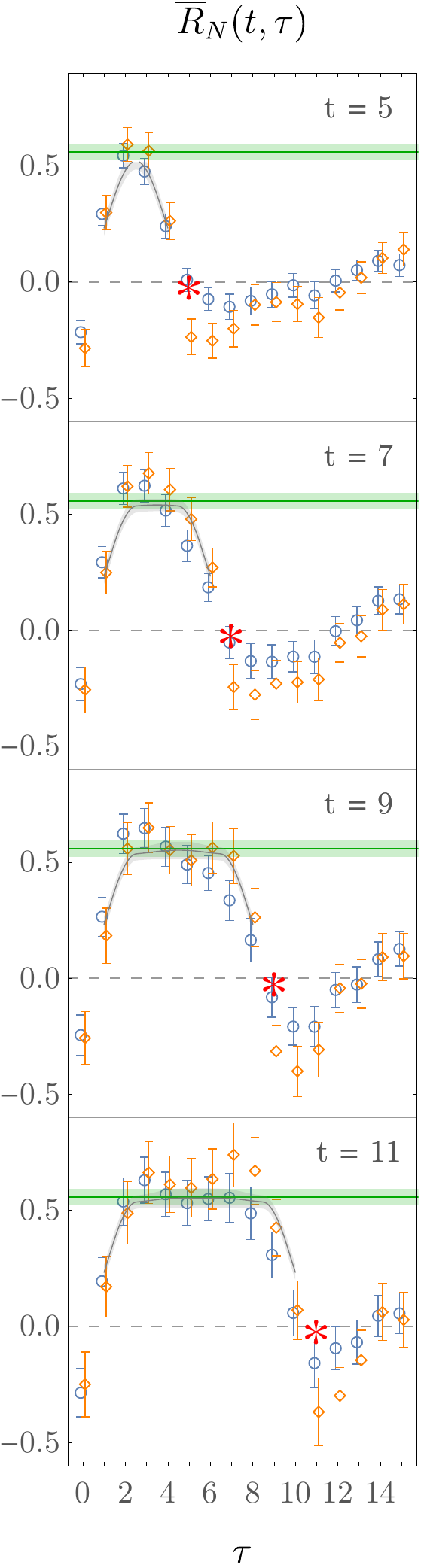}
	\includegraphics[height=0.68\textheight]{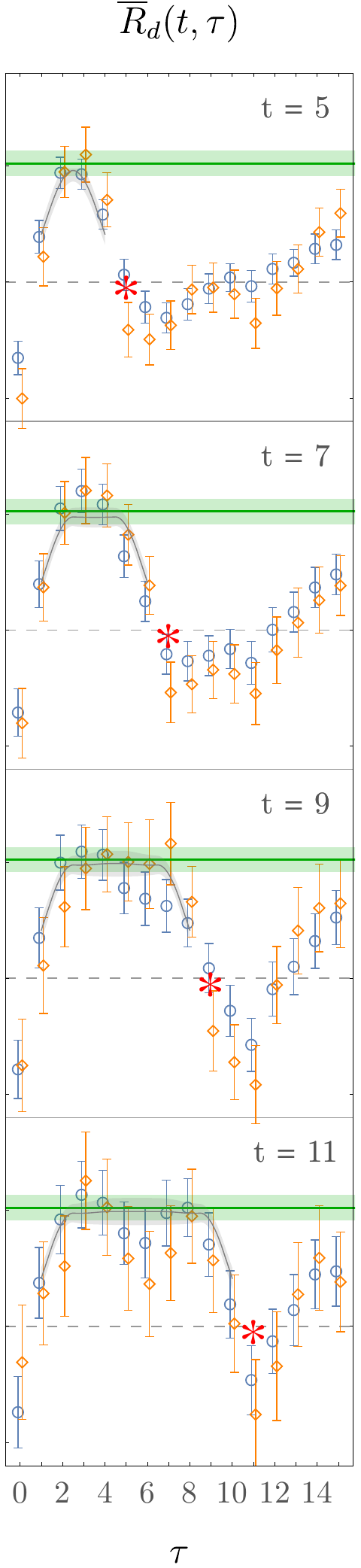}
	\includegraphics[height=0.68\textheight]{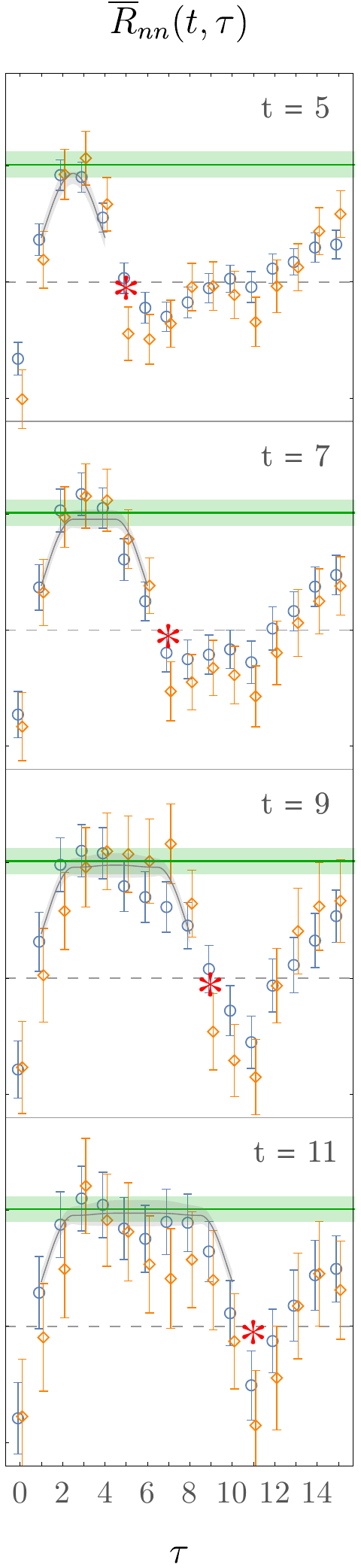}
	\includegraphics[height=0.68\textheight]{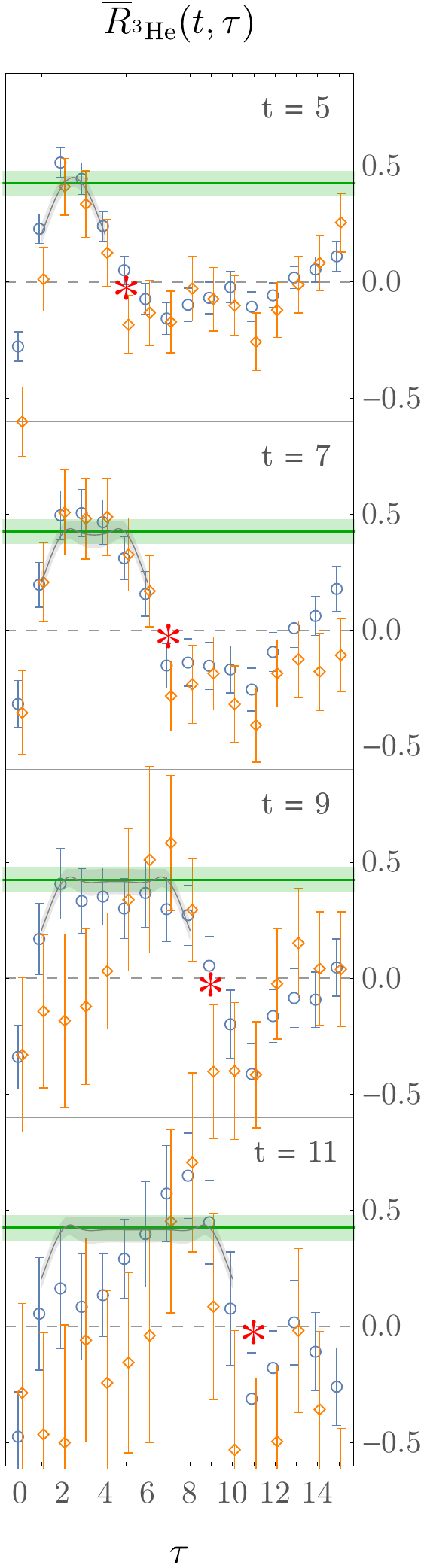}
	\caption{\label{fig:bumps800}Fits to the ratios $\overline{R}_h$ for $h$ = \{$N$, $d$, $nn$, $^3$He\}, averaged over operators in the $\tau_1^{(3)}$ irreducible representation and for different momenta with appropriate factors through Eqs.~\eqref{eq:Jhalf} and \eqref{eq:Jone} to extract $b_2^{(h)}$, for Ensemble B (with $m_\pi \sim 806$ MeV). Blue circles and orange diamonds denote SP and SS source-sink smearing combinations, respectively, while the red stars denote the sink times. The green bands (repeated on each panel) show the extracted results for the matrix elements from fits in the $t$--$\tau$ plane to the SP ratios using the exponential form of Eq.~\eqref{eq:rat}. The grey curves show those exponential fits. The scales on each figure are the same, and the light grey lines indicate the zero of each panel.}
\end{figure}

The averaged ratios $\overline{R}_h$ constructed in this way are fit using the exponential form in $t$ and $\tau$ derived in Eq.~\eqref{eq:rat}, which explicitly accounts for contributions from the first excited state in both the three and two-point functions (note that using the full two-state forms of Eqs.~\eqref{eq:C2pt} and \eqref{eq:C3pt}, rather than the expansion of Eq.~\eqref{eq:rat}, yields equivalent results).
A systematic uncertainty, determined by considering the variation of the extracted results over all choices of fit range with $\chi^2$/d.o.f $< \chi^2_\text{min.}/\text{d.o.f}+1$, is combined in quadrature with the statistical uncertainty of the best-fit result. Fits are constrained to begin no earlier in $t$ or $\tau$ than the time at which the relevant two-point function is consistent with a two-exponential form indicating significant contributions from only a single excited state (note that this consistency condition implies a time window that can be longer than that used for the best fits shown in Section~\ref{sec:nuclei}). The energy differences between the ground and first excited states which are determined in this way have much larger uncertainties, but are broadly consistent with, those obtained from the fits to the two-point functions.
The fits used to extract the $b_2^{(h)}$ are illustrated in Figs.~\ref{fig:bumps450} and \ref{fig:bumps800} for Ensemble A (with $m_\pi \sim 450$ MeV) and Ensemble B (with $m_\pi \sim 806$ MeV), respectively.
The momentum fractions extracted from these fits are entirely consistent with those determined using constant fits to the two-dimensional plateau regions of the ratios in the $t$--$\tau$ plane, but with smaller uncertainties because of the statistically more precise earlier times that can be included given the fit function accounting for excited states.
Moreover, results obtained using operators from the $\tau_3^{(6)}$ irreducible representation defined in Sec.~\ref{sec:SIops} are consistent, albeit typically with significantly larger uncertainties as fewer contributions (with different spins/polarisations, boosts, and operators) give non-zero contributions. 

\begin{figure}
	\includegraphics[width=0.49\textwidth]{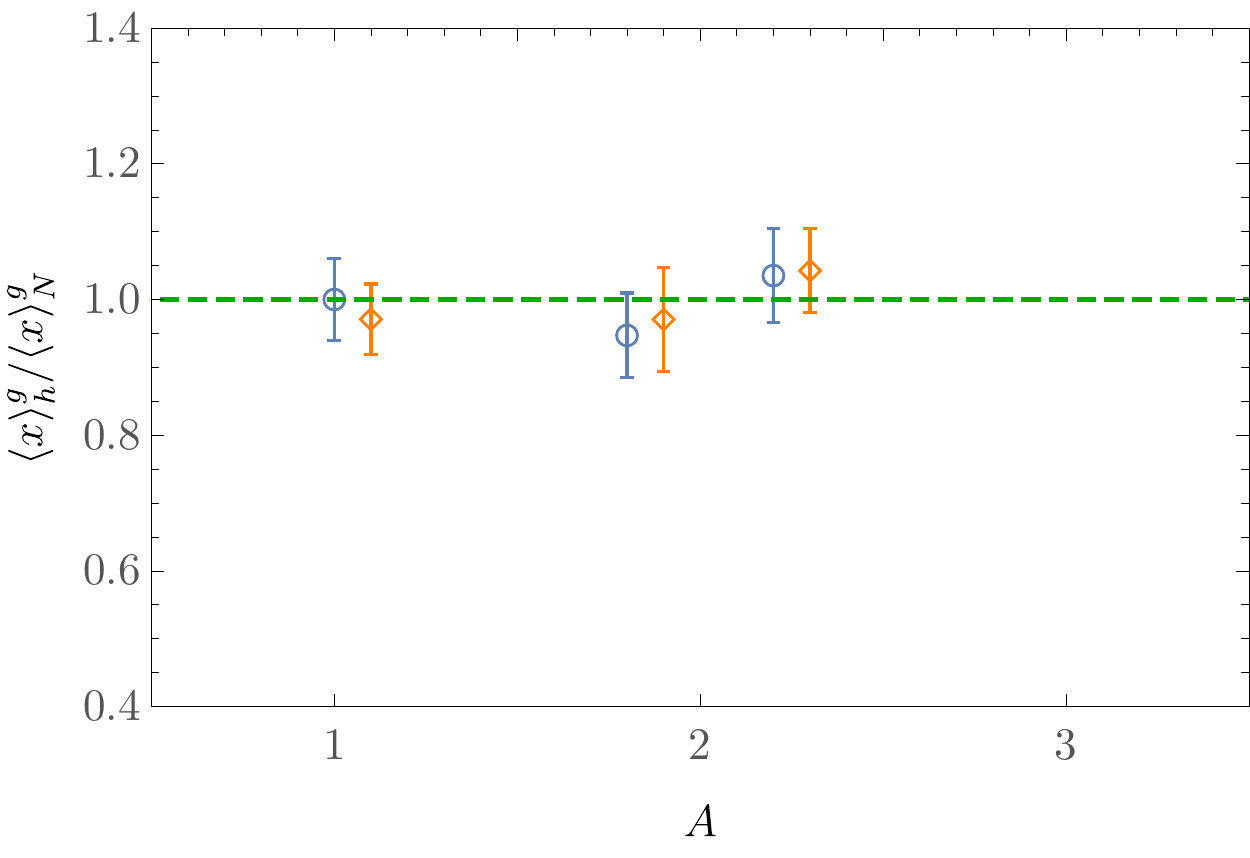}\hspace{2mm}
	\includegraphics[width=0.49\textwidth]{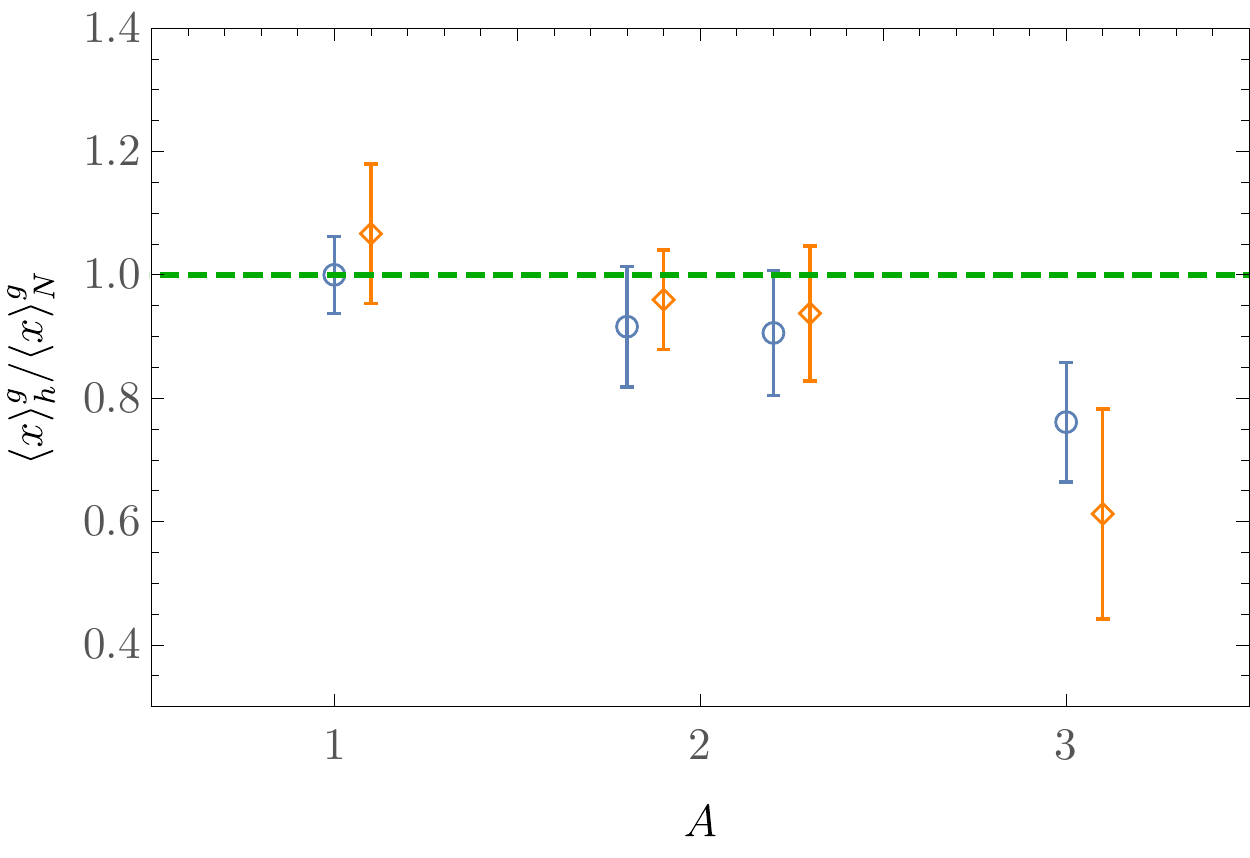}
	\caption{\label{fig:EMC}Ratios of the gluon momentum fraction in the nucleon and light nuclei $h$ = \{$N$, $d$, $nn$, $^3$He\} from left to right, to the central value of this quantity in the nucleon with SP source-sink smearing. The left and right panels show results for Ensemble A (with $m_\pi \sim 450$ MeV), and Ensemble B (with $m_\pi \sim 806$ MeV), respectively. The blue circles and orange diamonds denote results obtained using SP and SS source-sink smearing combinations. The green dashed line is at 1, shown to guide the eye. 
	}
\end{figure}

\begin{figure}
	\includegraphics[width=0.49\textwidth]{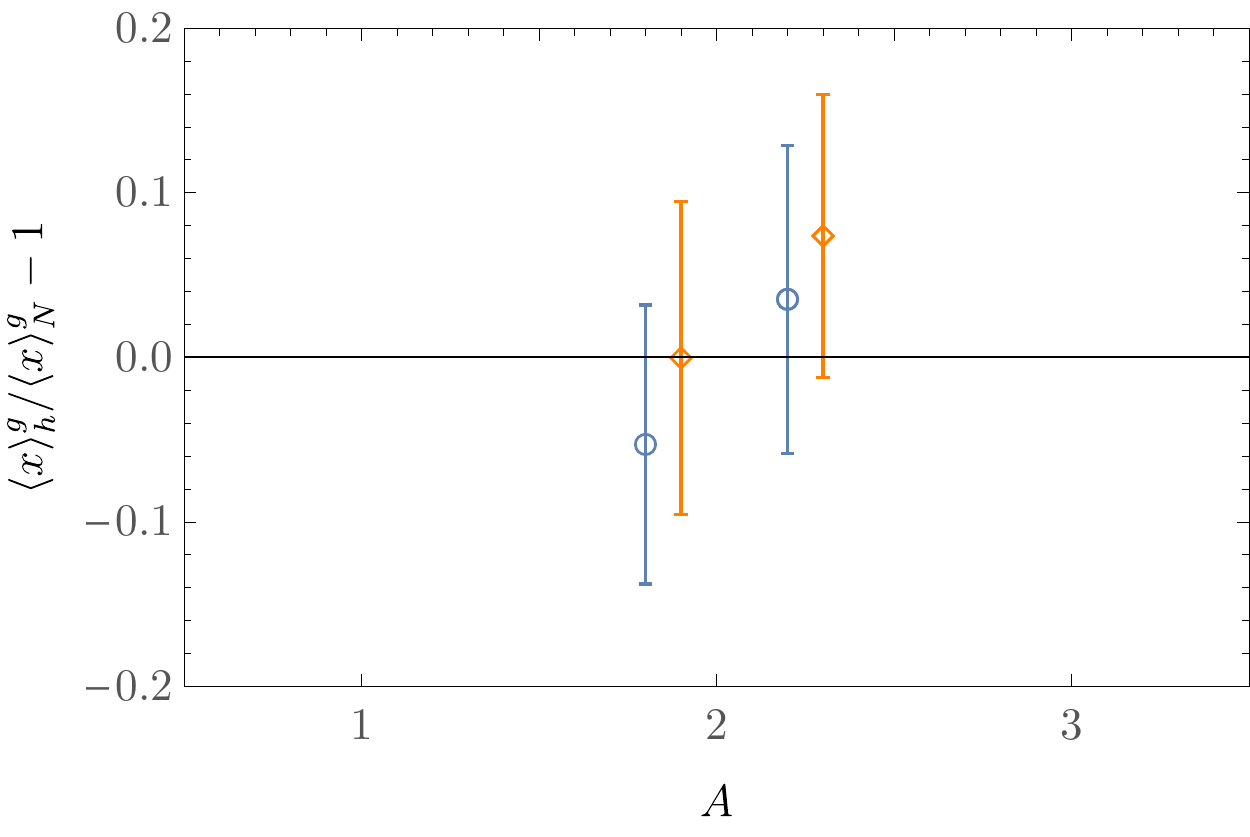}\hspace{2mm}
	\includegraphics[width=0.49\textwidth]{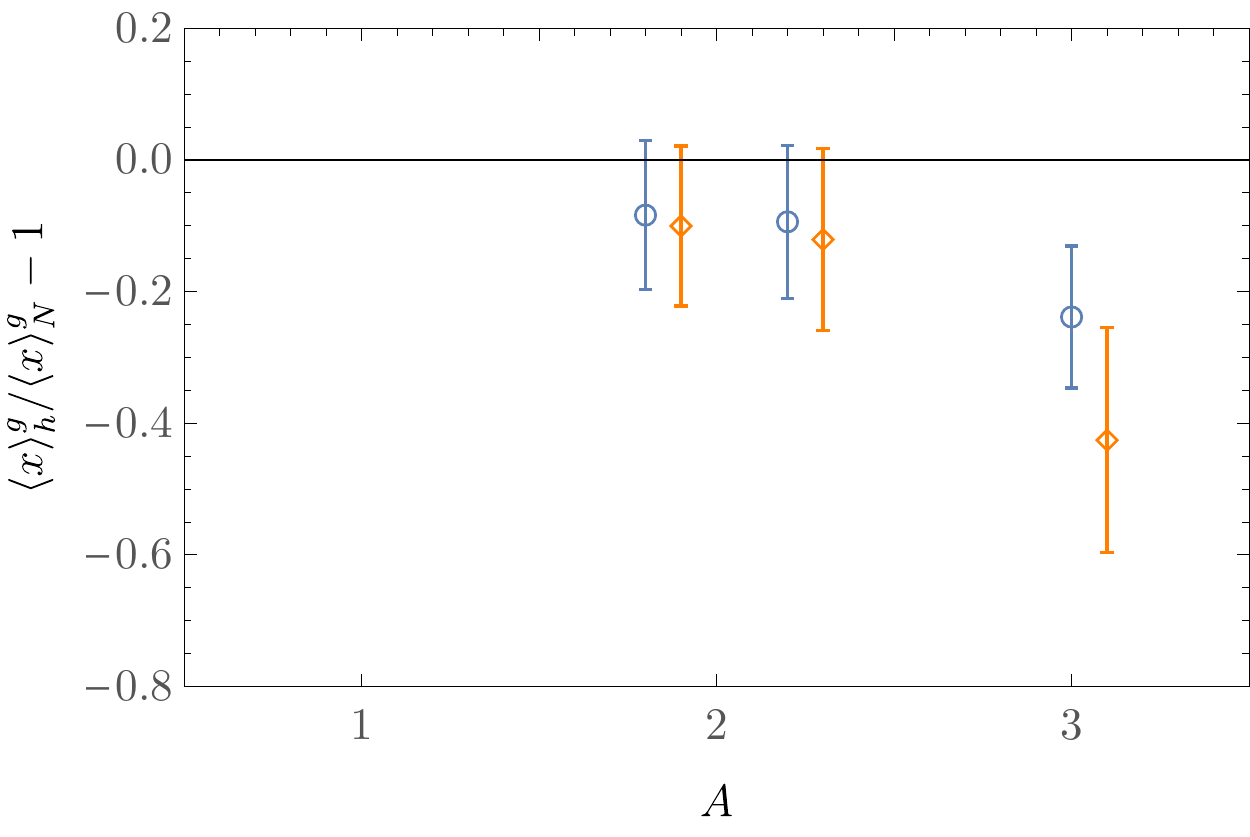}
	\caption{\label{fig:EMC2}Differences from unity of the ratios of the gluon momentum fraction in the nucleon and light nuclei $h$ = \{$N$, $d$, $nn$, $^3$He\} from left to right, to the nucleon with the same source-sink smearing. The left and right panels show results for Ensemble A (with $m_\pi \sim 450$ MeV), and Ensemble B (with $m_\pi \sim 806$ MeV), respectively. The blue circles and orange diamonds denote results obtained using SP and SS source-sink smearing combinations. 
	}
\end{figure}

The results, where SP and SS source-sink smearing combinations have been treated separately, are summarised in Figs.~\ref{fig:EMC} and \ref{fig:EMC2}, which shows the ratios of the gluon momentum fraction in the different light nuclei to that in the nucleon, for both Ensemble A and Ensemble B. As discussed in Sec.~\ref{sec:renorm}, the multiplicative renormalisation factors, which have not been calculated, cancel in these ratios. The effects of mixing with the isoscalar quark distributions are expected to be much smaller than the statistical uncertainties and give a negligible contribution to the uncertainties.
The resulting gluon momentum fractions of the light nuclei studied here are consistent with the gluon momentum fraction of the nucleon. For the $d$ and $nn$ states, this agreement is at the level of one standard deviation. The momentum fraction of $^3$He for Ensemble B (with $m_\pi \sim 806$~MeV), is consistent with that of the nucleon within two standard deviations.
An EMC effect in the gluon momentum fractions is thus found to be $< 10\%$ in the $A=2$ states at both unphysical sets of quark masses studied, and $< 25 \%$ in the $A=3$ state for $m_\pi \sim 806$~MeV.
Correlated differences of ratios do not provide
more precise constraints than separate fits, as can be seen by comparing Figs.~\ref{fig:EMC} and \ref{fig:EMC2}.

The calculations presented here use larger-than-physical values of the quark masses, where a gluonic analogue of the EMC effect is likely different from that in nature. Nevertheless, the results are consistent with expectations for the magnitude of such an effect in the gluon momentum fractions determined from phenomenological parameterisations of the quark and gluon distributions. In most cases, such parameterisations assume no nuclear effects in the deuteron. The difference between the gluon momentum fraction in $^3$He and the nucleon is poorly constrained, but is typically a positive shift of less than $5\%$ (e.g., using the CT14NNLO proton PDF~\cite{Dulat:2015mca} and the EPSS16~\cite{Eskola:2016oht} nuclear PDF sets).
As momentum is conserved, the difference between the quark momentum fractions in nuclei and in the proton also bounds the gluon EMC effect. This quantity is similarly typically less than $5\%$ for light nuclei in various PDF parameterisations~\cite{Dulat:2015mca,Eskola:2016oht,Kovarik:2015cma}. 

\begin{figure}
	\includegraphics[width=0.5\columnwidth]{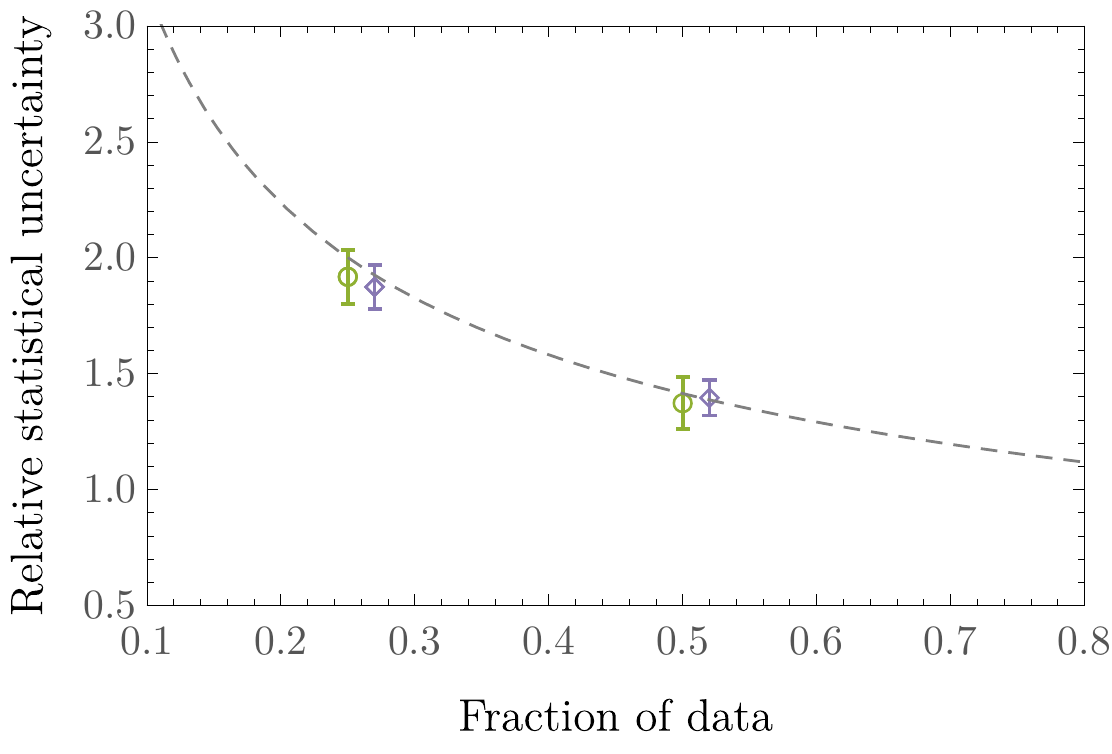}\qquad
	\caption{\label{fig:stat_scaling}Statistical scaling of the uncertainties of the correlation functions calculated on Ensemble A, with $m_\pi \sim 450$~MeV. The green circles and purple diamonds represent the scaling with number of configurations and number of sources per configuration, respectively. The grey dashed line corresponds to $\sqrt{N}$ statistical scaling.}
\end{figure}

The requirements for a future calculation to resolve a percent-level EMC effect as suggested by phenomenology can be estimated from the present calculations.
The scaling of the uncertainties of the correlation functions with statistics is shown in Fig.~\ref{fig:stat_scaling} for Ensemble A, which has the lighter pion mass of the two ensembles considered. Within uncertainties, no deviation from $\sqrt{N}$ statistical scaling is seen with either the number of configurations used or the number of sources used for propagator inversions on each configuration. 
Given this scaling behaviour, if the EMC effect with $m_\pi \sim 450$ MeV is of the same magnitude as the expectation from phenomenology, an order of magnitude more statistics than used in the present work will be required to cleanly resolve this effect. This would correspond to a few-percent determination of the proton and nuclear momentum fractions at this value of the pion masses.
It is possible that higher moments of the gluonic PDFs of nuclei, which are not subject to momentum conservation constraints, may have larger deviations from the nucleon moments. Since these correspond to gluon operators with additional derivatives they are statistically more difficult to resolve and no statistically significant signal was found on the ensembles studied in this work.

\begin{figure}
	\includegraphics[width=0.45\textwidth]{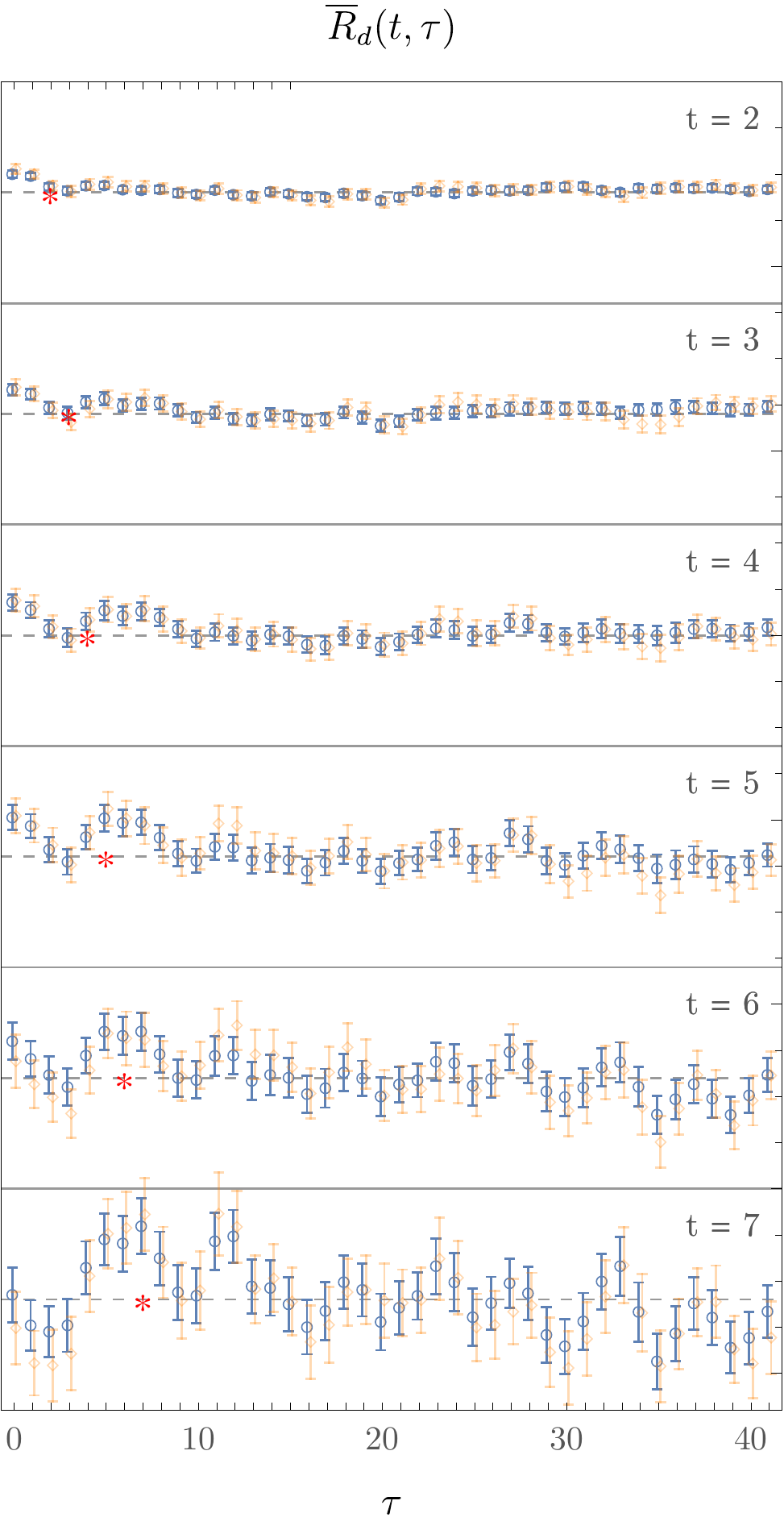}\hspace{2mm}
	\includegraphics[width=0.45\textwidth]{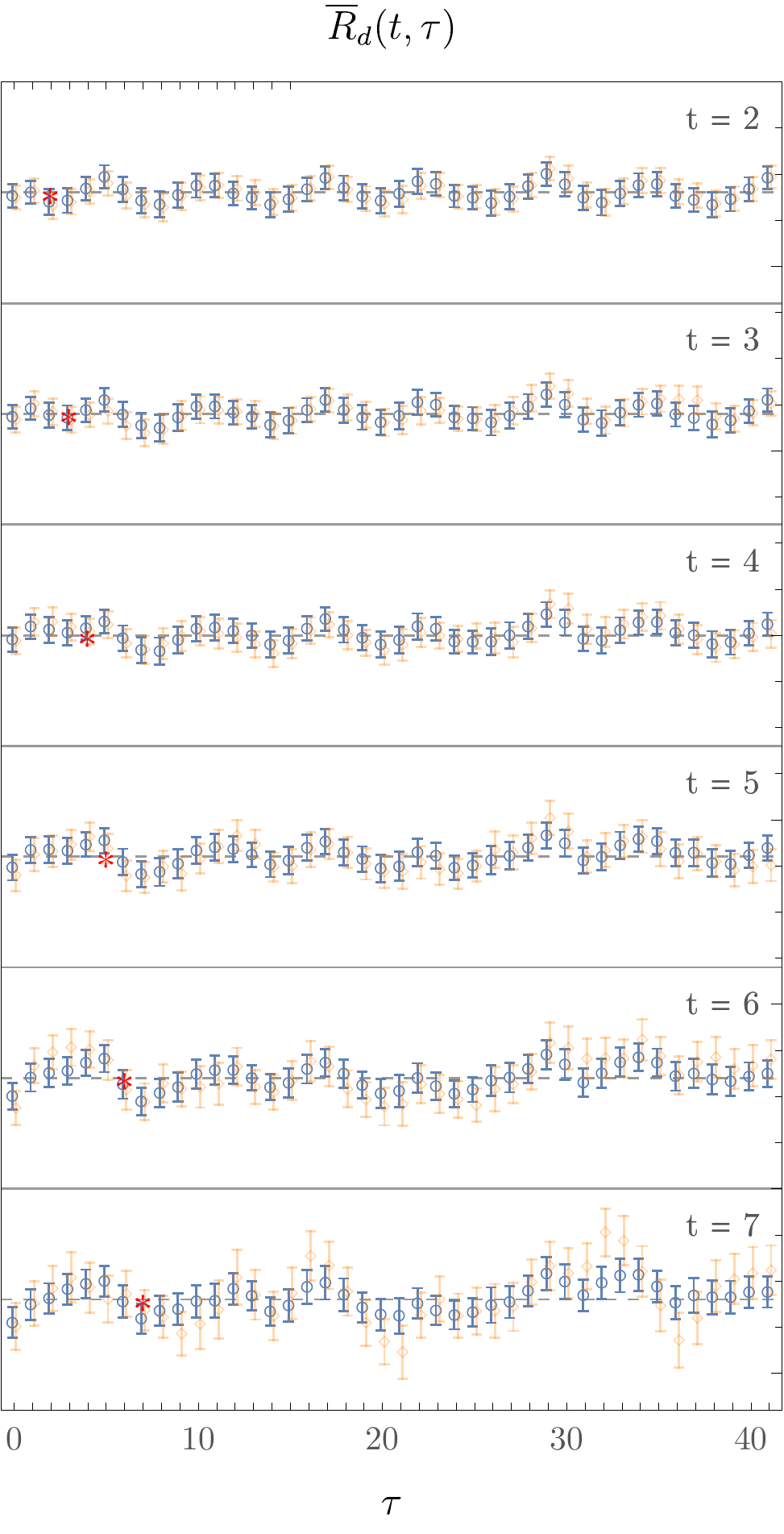}
	\caption{\label{fig:C2}The ratios $\overline{R}_d$ for the deuteron, averaged over operators in the $\tau_1^{(3)}$ irreducible representation and for different momenta with appropriate factors through Eq.~\eqref{eq:Jone} to extract $c_2^{(d)}$, for Ensembles A, with $m_\pi \sim 450$ MeV (left panel), and B, with $m_\pi \sim 806$ MeV (right panel). Blue circles and orange diamonds denote SP and SS source-sink smearing combinations, respectively, while the red stars denote the sink times. 
	}
\end{figure}

For the deuteron, the unpolarised gluon PDFs in the $j_z=\pm1$ and $j_z=0$ spin states are not necessarily the same, and the first Mellin moment of the difference is determined by $c_2^{(d)}$ in Eq.~\eqref{eq:Jone}. Through appropriate combinations of the averaged ratios $\overline{R}_d$ for different polarisations and momenta, $c_2^{(d)}$ can in principle be extracted from these calculations. However, at the current statistical precision, $c_2^{(d)}$ is not resolvable from zero for either ensemble, as shown in Fig.~\ref{fig:C2}. Taking the size of the uncertainties at $t=7$ as a bound on the size of the unrenormalised $c_2^{(d)}$ , it is apparent that $c_2^{(d)}/b_2^{(d)} \lsim 1/20$. This suppression is a natural consequence of the large $N_c$ scaling of $I\ne J$ operators~\cite{Kaplan:1995yg,Mattis:1988hg,Dai:1995zg,Dashen:1994qi,Dashen:1993jt} and the somewhat unnatural loosely-bound structure of the deuteron~\cite{Beane:2013br}.  

\section{Lowest Moment of Gluon Transversity in the Deuteron}
\label{sec:gt}

\begin{figure}
	\includegraphics[width=0.45\textwidth]{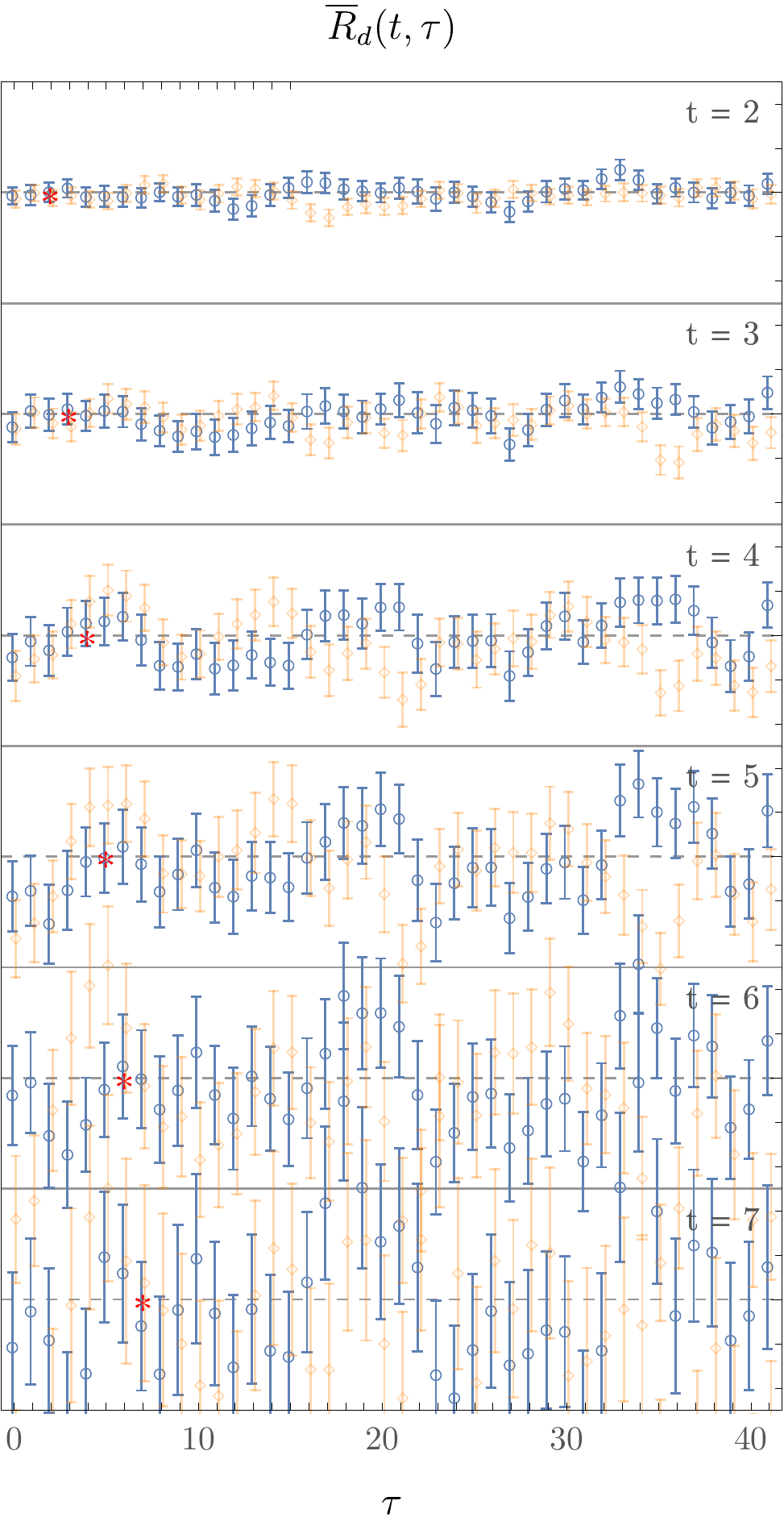}\hspace{2mm}
	\includegraphics[width=0.45\textwidth]{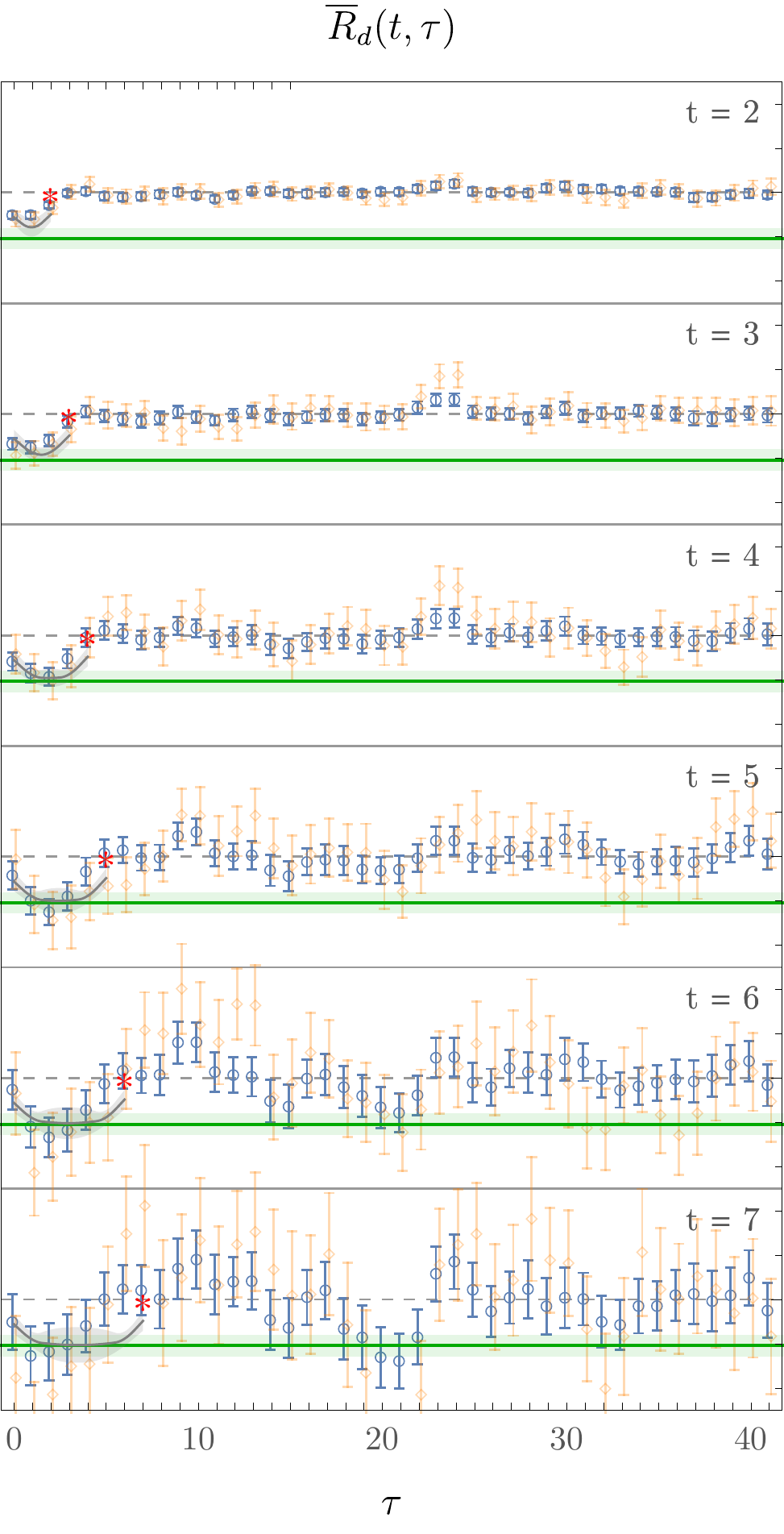}
	\caption{\label{fig:bumpsDG}Fits to the ratio $\overline{R}_d$, averaged over operators in the $\tau_1^{(2)}$ irreducible representation for the gluon transversity operator and for different momenta with appropriate factors through Eq.~\eqref{eq:JoneDG} to extract $a_2^{(d)}$, for Ensembles A with $m_\pi \sim 450$ MeV (left panel) and B, with $m_\pi \sim 806$ MeV (right panel). Blue circles and orange diamonds denote SP and SS source-sink smearing combinations, respectively, while the red star indicates the sink time $t$. For Ensemble B, the green bands (repeated on each panel) show the extracted results for the matrix elements from fits in the $t$--$\tau$ plane to the SP ratios using the exponential form of Eq.~\eqref{eq:rat}. The grey curves show those exponential fits. The scales on each figure (at both masses) are the same, and the light grey lines indicate the zero of each panel. A long time extent is shown to allow the scale of fluctuations in the noise region to be compared with the size of the signal.}
\end{figure}

The gluon transversity distribution is a direct measure of non-nucleonic gluon degrees of freedom in nuclei with spin $J\ge 1$~\cite{Jaffe:1989xy}. For this reason, it is a key target for several experimental programs~\cite{JLAB.LOI,Accardi:2012qut}, and a prediction of its magnitude in a light nucleus such as the deuteron would be very valuable in informing the design requirements of a potential EIC~\cite{Accardi:2012qut}. 
The deuteron matrix elements of the $n=2$ gluon transversity operator defined in Eq.~\eqref{eq:Jone} have been computed using the same procedure described above for the spin-independent gluon operator. At the current level of statistical precision a non-zero signal is observed in calculations with $m_\pi \sim 806$ MeV, while no signal can be resolved at $m_\pi \sim 450$ MeV. Fits to the relevant ratios of three and two-point functions, averaged over operators in the $\tau_1^{(2)}$ irreducible representation and for different momenta with appropriate factors as in Eq.~\eqref{eq:JoneDG}, are shown in Fig.~\ref{fig:bumpsDG}. The clear signal observed at $m_\pi \sim 806$ MeV, which is statistically non-zero to three standard deviations, represents the first evidence for non-nucleonic gluon contributions to nuclear structure, albeit at a larger-than-physical value of the light quark mass.

The unrenormalised result for the reduced matrix element of the $n=2$ gluon transversity operator at $m_\pi \sim 806$ MeV, defined in Eq.~\eqref{eq:JoneDG},  is $a_2^{(d)} = -0.010(3)$, obtained using a multi-exponential fit of the form of Eq.~\eqref{eq:multiexp} to the averaged ratio of three and two-point functions. This form is justified after $t=3$ where the deuteron two-point function is consistent with a two-exponential form (see Fig.~\ref{fig:effmassB}). Nevertheless, because of the degradation of the signal for this matrix element at sink times greater than $t\sim 6$ lattice units, necessitating a short fit window, the quoted bare result may still suffer from contamination from excited states that is not fully quantified.
This number can be compared with the unrenormalised result for the gluon momentum fraction in the deuteron, subject to these caveats and under the assumption that the multiplicative renormalisation of the two operators is similar. At $m_\pi \sim 806$ MeV, the bare result is $\langle x \rangle_g^{(d)}=b_2^{(d)}=0.51(5)$, giving a ratio of the transversity to spin-independent gluon structure quantities that is approximately an order of magnitude smaller than seen in previous studies of the $\phi$ meson~\cite{Detmold:2016gpy,Detmold:2017oqb} at $m_\pi \sim 450$~MeV (again, subject to assumptions about renormalisation effects cancelling in this ratio). This scaling is consistent with expectations from large-$N_c$, which predicts an additional suppression of $1/N_c^2$ of the transversity operator relative to the spin-independent operator in the deuteron~\cite{Kaplan:1995yg,Mattis:1988hg,Dai:1995zg,Dashen:1994qi,Dashen:1993jt} as compared to the ratio in a typical compact spin-1 hadron such as the $\phi$ meson.

Given the statistical scaling of the correlation functions at $m_\pi \sim 450$ MeV, illustrated in Fig.~\ref{fig:stat_scaling}, a factor of 5-10 improvement in statistics will be necessary to resolve the first moment of the gluon transversity structure function on this ensemble, if the magnitude of this quantity is of the same order at $m_\pi \sim 450$ MeV and $m_\pi \sim 806$ MeV. As discussed in the previous section, future calculations at that scale will also allow a few-percent determination of the proton and nuclear momentum fractions at this value of the pion masses, and resolve the gluon EMC effect if it has a similar size at $m_\pi \sim 450$~MeV as predicted from phenomenology. 

\section{Discussion}
\label{sec:discussion}

Since gluons do not carry electric charge, their distributions in hadrons and nuclei are considerably more difficult to probe than the corresponding quark distributions. Achieving a better understanding of this gluon structure is the mission driving a number of proposed experimental programs including the planned EIC, which will have the ability to measure gluonic properties of nuclei with unprecedented precision.
In particular, an analogue of the  EMC effect for gluons, in which the distributions of gluons in a nucleus differ from the sum of the gluon distributions in the constituent protons and neutrons, has not been observed but would be a milestone discovery of the EIC. 

The present study demonstrates that QCD predictions for gluonic structure quantities in nuclei can be obtained using LQCD. 
While no signal is seen for the first Mellin moment of the difference between the $j_x=\pm1$ and $j_z=0$ deuteron states, the unpolarised gluon longitudinal momentum fraction has been calculated in the nucleon and in light nuclei with $A\le 2$ with 10\% uncertainties at quark masses corresponding to $m_{\pi} \sim $~450 and 806~MeV, and A = 3 with 25\% uncertainty for $m_{\pi} \sim 806$~MeV. 
The gluon momentum fractions in the spin-averaged nuclei are broadly consistent with that in the nucleon, indicating that a gluonic EMC-like effect in this quantity is smaller than these bounds.
With an order of magnitude more statistics, these calculations could discern a few-percent level effect in this quantity, as suggested by phenomenology.
Higher moments of gluon distributions and other partonic quantities will also be able to be resolved at these statistics. With lighter values of the quark masses, these calculations will be able to be extrapolated to the physical quark masses using effective field theory, see Refs.~\cite{Chen:2004zx,Beane:2004xf,Chen:2016bde}.
Such future LQCD studies with fully-controlled uncertainties will be influential in the development and planning of the EIC by providing a precision target for this key quantity.

The first moment of the gluonic transversity distribution in the deuteron has also been investigated. This quantity is particularly interesting since it is a clean measure of gluonic degrees of freedom, safe from leading-twist mixing with quark distributions. Moreover, in nuclei with spin $J\ge1$, it has been recognised~\cite{Jaffe:1989xy} that the corresponding structure function is sensitive to `exotic gluons' as neither nucleons nor pions inside a nucleus (nor anything with spin less than one) can transfer two units of helicity to the nuclear target. 
A clear signal for the gluonic transversity was observed at $m_\pi \sim 806$ MeV. The value indicates a suppression of this quantity relative to the momentum fraction that is a factor of 5-10 more severe than observed in LQCD calculations of the same distributions in a $\phi$ meson~\cite{Detmold:2016gpy,Detmold:2017oqb}, as predicted by large-$N_c$~\cite{Kaplan:1995yg,Mattis:1988hg,Dai:1995zg,Dashen:1994qi,Dashen:1993jt}.
Future calculations of this quantity at lighter values of the pion mass will allow this result to be extrapolated to the physical point and will provide important information guiding design requirements for experiments at Jefferson Lab~\cite{JLAB.LOI} and the EIC that seek to perform first measurements of non-nucleonic gluons in nuclei.

\begin{acknowledgments}
	We thank Silas Beane, Emmanuel Chang, Zohreh Davoudi, Bob Jaffe, James Maxwell, Richard Milner, Assumpta Parre\~no, and Brian Tiburzi for helpful discussions.   
	Calculations were performed using computational resources provided by NERSC (supported by U.S. Department of Energy grant number DE-AC02-05CH11231), and by the USQCD collaboration through LQCD Ext-II Hardware Project. This research used resources of the Oak Ridge Leadership Computing Facility at the Oak Ridge National Laboratory, which is supported by the Office of Science of the U.S. Department of Energy under Contract number DE-AC05-00OR22725. 
	Parts of the calculations used the {\tt chroma} software suite~\cite{Edwards:2004sx} and the {\tt quda} library~\cite{Clark:2009wm,Babich:2011np}. 
	FW was partially supported through the USQCD Scientific Discovery through Advanced Computing (SciDAC) project funded by U.S. Department of Energy, Office of Science, Offices of Advanced Scientific Computing Research, Nuclear Physics and High Energy Physics and by the U.S. Department of Energy, Office of Science, Office of Nuclear Physics under contract
	DE-AC05-06OR23177.
	The work of WD is supported in part by the U.S. Department of Energy, Office of Science, Office of Nuclear Physics, within the framework of the TMD Topical Collaboration. 
	WD and PES were partly supported by U.S. Department of Energy Early Career Research Award DE-SC0010495 and grant number DE-SC0011090. KO was partially supported by the U.S. Department of Energy through grant number DE-FG02-04ER41302 and PES and KO were partially supported through contract number DE-AC05-06OR23177 under which JSA operates the Thomas Jefferson National Accelerator Facility. MJS was supported by DOE grant number DE-FG02-00ER41132. MLW was supported in part by DOE grant number DE-FG02-00ER41132 and also by an MIT Pappalardo Fellowship. 
\end{acknowledgments}

\bibliography{nuclearglue}

\begin{thebibliography}{75}
\expandafter\ifx\csname natexlab\endcsname\relax\def\natexlab#1{#1}\fi
\expandafter\ifx\csname bibnamefont\endcsname\relax
  \def\bibnamefont#1{#1}\fi
\expandafter\ifx\csname bibfnamefont\endcsname\relax
  \def\bibfnamefont#1{#1}\fi
\expandafter\ifx\csname citenamefont\endcsname\relax
  \def\citenamefont#1{#1}\fi
\expandafter\ifx\csname url\endcsname\relax
  \def\url#1{\texttt{#1}}\fi
\expandafter\ifx\csname urlprefix\endcsname\relax\def\urlprefix{URL }\fi
\providecommand{\bibinfo}[2]{#2}
\providecommand{\eprint}[2][]{\url{#2}}

\bibitem[{\citenamefont{Dudek et~al.}(2012)}]{Dudek:2012vr}
\bibinfo{author}{\bibfnamefont{J.}~\bibnamefont{Dudek}} \bibnamefont{et~al.},
  \bibinfo{journal}{Eur. Phys. J.} \textbf{\bibinfo{volume}{A48}},
  \bibinfo{pages}{187} (\bibinfo{year}{2012}).

\bibitem[{\citenamefont{Armstrong
  et~al.}(2017{\natexlab{a}})}]{Armstrong:2017wfw}
\bibinfo{author}{\bibfnamefont{W.}~\bibnamefont{Armstrong}}
  \bibnamefont{et~al.} (\bibinfo{year}{2017}{\natexlab{a}}),
  \eprint{1708.00888}.

\bibitem[{\citenamefont{Armstrong
  et~al.}(2017{\natexlab{b}})}]{Armstrong:2017zqr}
\bibinfo{author}{\bibfnamefont{W.}~\bibnamefont{Armstrong}}
  \bibnamefont{et~al.} (\bibinfo{year}{2017}{\natexlab{b}}),
  \eprint{1708.00891}.

\bibitem[{\citenamefont{Abbon et~al.}(2007)}]{Abbon:2007pq}
\bibinfo{author}{\bibfnamefont{P.}~\bibnamefont{Abbon}} \bibnamefont{et~al.}
  (\bibinfo{collaboration}{COMPASS}), \bibinfo{journal}{Nucl. Instrum. Meth.}
  \textbf{\bibinfo{volume}{A577}}, \bibinfo{pages}{455} (\bibinfo{year}{2007}).

\bibitem[{\citenamefont{Aschenauer et~al.}(2015)}]{Aschenauer:2015eha}
\bibinfo{author}{\bibfnamefont{E.-C.} \bibnamefont{Aschenauer}}
  \bibnamefont{et~al.} (\bibinfo{year}{2015}).

\bibitem[{\citenamefont{Aubert et~al.}(1983)}]{Aubert:1983xm}
\bibinfo{author}{\bibfnamefont{J.~J.} \bibnamefont{Aubert}}
  \bibnamefont{et~al.} (\bibinfo{collaboration}{European Muon}),
  \bibinfo{journal}{Phys. Lett.} \textbf{\bibinfo{volume}{B123}},
  \bibinfo{pages}{275} (\bibinfo{year}{1983}).

\bibitem[{\citenamefont{Norton}(2003)}]{Norton:2003cb}
\bibinfo{author}{\bibfnamefont{P.~R.} \bibnamefont{Norton}},
  \bibinfo{journal}{Rept. Prog. Phys.} \textbf{\bibinfo{volume}{66}},
  \bibinfo{pages}{1253} (\bibinfo{year}{2003}).

\bibitem[{\citenamefont{Geesaman et~al.}(1995)\citenamefont{Geesaman, Saito,
  and Thomas}}]{Geesaman:1995yd}
\bibinfo{author}{\bibfnamefont{D.~F.} \bibnamefont{Geesaman}},
  \bibinfo{author}{\bibfnamefont{K.}~\bibnamefont{Saito}}, \bibnamefont{and}
  \bibinfo{author}{\bibfnamefont{A.~W.} \bibnamefont{Thomas}},
  \bibinfo{journal}{Ann. Rev. Nucl. Part. Sci.} \textbf{\bibinfo{volume}{45}},
  \bibinfo{pages}{337} (\bibinfo{year}{1995}).

\bibitem[{\citenamefont{{Malace, Simona and Gaskell, David and Higinbotham,
  Douglas W. and Clo\"et, Ian}}(2014)}]{Malace:2014uea}
\bibinfo{author}{\bibnamefont{{Malace, Simona and Gaskell, David and
  Higinbotham, Douglas W. and Clo\"et, Ian}}}, \bibinfo{journal}{Int. J. Mod.
  Phys.} \textbf{\bibinfo{volume}{E23}}, \bibinfo{pages}{1430013}
  (\bibinfo{year}{2014}), \eprint{1405.1270}.

\bibitem[{\citenamefont{Arneodo}(1994)}]{Arneodo:1992wf}
\bibinfo{author}{\bibfnamefont{M.}~\bibnamefont{Arneodo}},
  \bibinfo{journal}{Phys. Rept.} \textbf{\bibinfo{volume}{240}},
  \bibinfo{pages}{301} (\bibinfo{year}{1994}).

\bibitem[{\citenamefont{Piller and Weise}(2000)}]{Piller:1999wx}
\bibinfo{author}{\bibfnamefont{G.}~\bibnamefont{Piller}} \bibnamefont{and}
  \bibinfo{author}{\bibfnamefont{W.}~\bibnamefont{Weise}},
  \bibinfo{journal}{Phys. Rept.} \textbf{\bibinfo{volume}{330}},
  \bibinfo{pages}{1} (\bibinfo{year}{2000}).

\bibitem[{\citenamefont{Hen et~al.}(2013)\citenamefont{Hen, Higinbotham,
  Miller, Piasetzky, and Weinstein}}]{Hen:2013oha}
\bibinfo{author}{\bibfnamefont{O.}~\bibnamefont{Hen}},
  \bibinfo{author}{\bibfnamefont{D.~W.} \bibnamefont{Higinbotham}},
  \bibinfo{author}{\bibfnamefont{G.~A.} \bibnamefont{Miller}},
  \bibinfo{author}{\bibfnamefont{E.}~\bibnamefont{Piasetzky}},
  \bibnamefont{and} \bibinfo{author}{\bibfnamefont{L.~B.}
  \bibnamefont{Weinstein}}, \bibinfo{journal}{Int. J. Mod. Phys. E}
  \textbf{\bibinfo{volume}{22}}, \bibinfo{pages}{1330017}
  (\bibinfo{year}{2013}).

\bibitem[{\citenamefont{Baglio and Djouadi}(2011)}]{Baglio:2010ae}
\bibinfo{author}{\bibfnamefont{J.}~\bibnamefont{Baglio}} \bibnamefont{and}
  \bibinfo{author}{\bibfnamefont{A.}~\bibnamefont{Djouadi}},
  \bibinfo{journal}{JHEP} \textbf{\bibinfo{volume}{03}}, \bibinfo{pages}{055}
  (\bibinfo{year}{2011}), \eprint{1012.0530}.

\bibitem[{\citenamefont{Nakagawa}(2017)}]{Nakagawa:2017ljf}
\bibinfo{author}{\bibfnamefont{I.}~\bibnamefont{Nakagawa}}
  (\bibinfo{collaboration}{PHENIX}), \bibinfo{journal}{EPJ Web Conf.}
  \textbf{\bibinfo{volume}{141}}, \bibinfo{pages}{03006}
  (\bibinfo{year}{2017}).

\bibitem[{\citenamefont{Franco}(2016)}]{Franco:2016aet}
\bibinfo{author}{\bibfnamefont{C.}~\bibnamefont{Franco}}
  (\bibinfo{collaboration}{COMPASS}), \bibinfo{journal}{Nucl. Part. Phys.
  Proc.} \textbf{\bibinfo{volume}{273-275}}, \bibinfo{pages}{2084}
  (\bibinfo{year}{2016}).

\bibitem[{\citenamefont{Jaffe and Manohar}(1989)}]{Jaffe:1989xy}
\bibinfo{author}{\bibfnamefont{R.~L.} \bibnamefont{Jaffe}} \bibnamefont{and}
  \bibinfo{author}{\bibfnamefont{A.}~\bibnamefont{Manohar}},
  \bibinfo{journal}{Phys. Lett.} \textbf{\bibinfo{volume}{B223}},
  \bibinfo{pages}{218} (\bibinfo{year}{1989}).

\bibitem[{\citenamefont{Hoodbhoy and Ji}(1998)}]{Hoodbhoy:1998vm}
\bibinfo{author}{\bibfnamefont{P.}~\bibnamefont{Hoodbhoy}} \bibnamefont{and}
  \bibinfo{author}{\bibfnamefont{X.-D.} \bibnamefont{Ji}},
  \bibinfo{journal}{Phys. Rev.} \textbf{\bibinfo{volume}{D58}},
  \bibinfo{pages}{054006} (\bibinfo{year}{1998}), \eprint{hep-ph/9801369}.

\bibitem[{\citenamefont{Diehl}(2003)}]{Diehl:2003ny}
\bibinfo{author}{\bibfnamefont{M.}~\bibnamefont{Diehl}},
  \bibinfo{journal}{Phys. Rept.} \textbf{\bibinfo{volume}{388}},
  \bibinfo{pages}{41} (\bibinfo{year}{2003}), \eprint{hep-ph/0307382}.

\bibitem[{\citenamefont{Berger et~al.}(2001)\citenamefont{Berger, Cano, Diehl,
  and Pire}}]{Berger:2001zb}
\bibinfo{author}{\bibfnamefont{E.~R.} \bibnamefont{Berger}},
  \bibinfo{author}{\bibfnamefont{F.}~\bibnamefont{Cano}},
  \bibinfo{author}{\bibfnamefont{M.}~\bibnamefont{Diehl}}, \bibnamefont{and}
  \bibinfo{author}{\bibfnamefont{B.}~\bibnamefont{Pire}},
  \bibinfo{journal}{Phys. Rev. Lett.} \textbf{\bibinfo{volume}{87}},
  \bibinfo{pages}{142302} (\bibinfo{year}{2001}), \eprint{hep-ph/0106192}.

\bibitem[{\citenamefont{Liuti and Kathuria}(2014)}]{Liuti:2014dda}
\bibinfo{author}{\bibfnamefont{S.}~\bibnamefont{Liuti}} \bibnamefont{and}
  \bibinfo{author}{\bibfnamefont{K.}~\bibnamefont{Kathuria}},
  \bibinfo{journal}{J. Phys. Conf. Ser.} \textbf{\bibinfo{volume}{543}},
  \bibinfo{pages}{012005} (\bibinfo{year}{2014}).

\bibitem[{\citenamefont{Mulders and Rodrigues}(2001)}]{Mulders:2000sh}
\bibinfo{author}{\bibfnamefont{P.~J.} \bibnamefont{Mulders}} \bibnamefont{and}
  \bibinfo{author}{\bibfnamefont{J.}~\bibnamefont{Rodrigues}},
  \bibinfo{journal}{Phys. Rev.} \textbf{\bibinfo{volume}{D63}},
  \bibinfo{pages}{094021} (\bibinfo{year}{2001}), \eprint{hep-ph/0009343}.

\bibitem[{\citenamefont{Aprahamian et~al.}(2015)\citenamefont{Aprahamian,
  Atcher, Caines, Cates, Cizewski, Cirigliano, Dean, Deshpande, Ent, Fahey
  et~al.}}]{Geesaman:2015fha}
\bibinfo{author}{\bibfnamefont{A.}~\bibnamefont{Aprahamian}},
  \bibinfo{author}{\bibfnamefont{R.}~\bibnamefont{Atcher}},
  \bibinfo{author}{\bibfnamefont{H.}~\bibnamefont{Caines}},
  \bibinfo{author}{\bibfnamefont{G.}~\bibnamefont{Cates}},
  \bibinfo{author}{\bibfnamefont{J.~A.} \bibnamefont{Cizewski}},
  \bibinfo{author}{\bibfnamefont{V.}~\bibnamefont{Cirigliano}},
  \bibinfo{author}{\bibfnamefont{D.~J.} \bibnamefont{Dean}},
  \bibinfo{author}{\bibfnamefont{A.}~\bibnamefont{Deshpande}},
  \bibinfo{author}{\bibfnamefont{R.}~\bibnamefont{Ent}},
  \bibinfo{author}{\bibfnamefont{F.}~\bibnamefont{Fahey}}, \bibnamefont{et~al.}
  (\bibinfo{year}{2015}).

\bibitem[{\citenamefont{A.~Accardi et~al.}(2012)\citenamefont{A.~Accardi,
  Anselmino, Armesto, Aschenauer, Bacchetta, Boer, Brooks, Burton, Chang, Deng
  et~al.}}]{Accardi:2012qut}
\bibinfo{author}{\bibfnamefont{J.~A.} \bibnamefont{A.~Accardi}},
  \bibinfo{author}{\bibfnamefont{M.}~\bibnamefont{Anselmino}},
  \bibinfo{author}{\bibfnamefont{N.}~\bibnamefont{Armesto}},
  \bibinfo{author}{\bibfnamefont{E.}~\bibnamefont{Aschenauer}},
  \bibinfo{author}{\bibfnamefont{A.}~\bibnamefont{Bacchetta}},
  \bibinfo{author}{\bibfnamefont{D.}~\bibnamefont{Boer}},
  \bibinfo{author}{\bibfnamefont{W.}~\bibnamefont{Brooks}},
  \bibinfo{author}{\bibfnamefont{T.}~\bibnamefont{Burton}},
  \bibinfo{author}{\bibfnamefont{N.-B.} \bibnamefont{Chang}},
  \bibinfo{author}{\bibfnamefont{W.-T.} \bibnamefont{Deng}},
  \bibnamefont{et~al.} (\bibinfo{year}{2012}), \eprint{1212.1701}.

\bibitem[{\citenamefont{Ent}(2016)}]{Ent:2016lod}
\bibinfo{author}{\bibfnamefont{R.}~\bibnamefont{Ent}}, \bibinfo{journal}{Eur.
  Phys. J.} \textbf{\bibinfo{volume}{A52}}, \bibinfo{pages}{162}
  (\bibinfo{year}{2016}).

\bibitem[{\citenamefont{Boer}(2016)}]{Boer:2016jnn}
\bibinfo{author}{\bibfnamefont{D.}~\bibnamefont{Boer}}, \bibinfo{journal}{EPJ
  Web Conf.} \textbf{\bibinfo{volume}{112}}, \bibinfo{pages}{02003}
  (\bibinfo{year}{2016}), \eprint{1601.01813}.

\bibitem[{\citenamefont{Pisano et~al.}(2017)\citenamefont{Pisano, Boer,
  Mulders, and Zhou}}]{Pisano:2016jjf}
\bibinfo{author}{\bibfnamefont{C.}~\bibnamefont{Pisano}},
  \bibinfo{author}{\bibfnamefont{D.}~\bibnamefont{Boer}},
  \bibinfo{author}{\bibfnamefont{P.~J.} \bibnamefont{Mulders}},
  \bibnamefont{and} \bibinfo{author}{\bibfnamefont{J.}~\bibnamefont{Zhou}},
  \bibinfo{journal}{PoS} \textbf{\bibinfo{volume}{QCDEV2016}},
  \bibinfo{pages}{028} (\bibinfo{year}{2017}), \eprint{1612.06611}.

\bibitem[{\citenamefont{Chudakov et~al.}(2016)\citenamefont{Chudakov,
  Higinbotham, Hyde, Furletov, Furletova, Nguyen, Stratmann, Strikman, Weiss,
  and Yoshida}}]{Chudakov:2016otl}
\bibinfo{author}{\bibfnamefont{E.}~\bibnamefont{Chudakov}},
  \bibinfo{author}{\bibfnamefont{D.}~\bibnamefont{Higinbotham}},
  \bibinfo{author}{\bibfnamefont{C.}~\bibnamefont{Hyde}},
  \bibinfo{author}{\bibfnamefont{S.}~\bibnamefont{Furletov}},
  \bibinfo{author}{\bibfnamefont{{\relax Yu}.}~\bibnamefont{Furletova}},
  \bibinfo{author}{\bibfnamefont{D.}~\bibnamefont{Nguyen}},
  \bibinfo{author}{\bibfnamefont{M.}~\bibnamefont{Stratmann}},
  \bibinfo{author}{\bibfnamefont{M.}~\bibnamefont{Strikman}},
  \bibinfo{author}{\bibfnamefont{C.}~\bibnamefont{Weiss}}, \bibnamefont{and}
  \bibinfo{author}{\bibfnamefont{R.}~\bibnamefont{Yoshida}},
  \bibinfo{journal}{PoS} \textbf{\bibinfo{volume}{DIS2016}},
  \bibinfo{pages}{143} (\bibinfo{year}{2016}), \eprint{1608.08686}.

\bibitem[{\citenamefont{Maxwell et~al.}(2014)\citenamefont{Maxwell, Detmold,
  Jaffe, Milner, Crabb, Day, Keller, Rondon, Jones, Keith et~al.}}]{JLAB.LOI}
\bibinfo{author}{\bibfnamefont{J.}~\bibnamefont{Maxwell}},
  \bibinfo{author}{\bibfnamefont{W.}~\bibnamefont{Detmold}},
  \bibinfo{author}{\bibfnamefont{R.}~\bibnamefont{Jaffe}},
  \bibinfo{author}{\bibfnamefont{R.}~\bibnamefont{Milner}},
  \bibinfo{author}{\bibfnamefont{D.}~\bibnamefont{Crabb}},
  \bibinfo{author}{\bibfnamefont{D.}~\bibnamefont{Day}},
  \bibinfo{author}{\bibfnamefont{D.}~\bibnamefont{Keller}},
  \bibinfo{author}{\bibfnamefont{O.~A.} \bibnamefont{Rondon}},
  \bibinfo{author}{\bibfnamefont{M.}~\bibnamefont{Jones}},
  \bibinfo{author}{\bibfnamefont{C.}~\bibnamefont{Keith}},
  \bibnamefont{et~al.}, \bibinfo{journal}{Jefferson Lab LOI12-14-001}
  (\bibinfo{year}{2014}).

\bibitem[{\citenamefont{Cosyn et~al.}(2016)\citenamefont{Cosyn, Guzey,
  Sargsian, Strikman, and Weiss}}]{Cosyn:2016oiq}
\bibinfo{author}{\bibfnamefont{W.}~\bibnamefont{Cosyn}},
  \bibinfo{author}{\bibfnamefont{V.}~\bibnamefont{Guzey}},
  \bibinfo{author}{\bibfnamefont{M.}~\bibnamefont{Sargsian}},
  \bibinfo{author}{\bibfnamefont{M.}~\bibnamefont{Strikman}}, \bibnamefont{and}
  \bibinfo{author}{\bibfnamefont{C.}~\bibnamefont{Weiss}},
  \bibinfo{journal}{EPJ Web Conf.} \textbf{\bibinfo{volume}{112}},
  \bibinfo{pages}{01022} (\bibinfo{year}{2016}), \eprint{1601.06665}.

\bibitem[{\citenamefont{{Engelhardt, M. and Musch, B. and H\"agler, P. and
  Negele, J. and Sch\"afer, A.}}(2015)}]{Engelhardt:2015jba}
\bibinfo{author}{\bibnamefont{{Engelhardt, M. and Musch, B. and H\"agler, P.
  and Negele, J. and Sch\"afer, A.}}}, \bibinfo{journal}{Few Body Syst.}
  \textbf{\bibinfo{volume}{56}}, \bibinfo{pages}{447} (\bibinfo{year}{2015}).

\bibitem[{\citenamefont{{Engelhardt, Michael and Musch, Bernhard and
  Bhattacharya, Tanmoy and Gupta, Rajan and H\"agler, Philipp and Krieg, Stefan
  and Negele, John and Pochinsky, Andrew and Syritsyn, Sergey and Yoon,
  Boram}}(2016)}]{Engelhardt:2015czw}
\bibinfo{author}{\bibnamefont{{Engelhardt, Michael and Musch, Bernhard and
  Bhattacharya, Tanmoy and Gupta, Rajan and H\"agler, Philipp and Krieg, Stefan
  and Negele, John and Pochinsky, Andrew and Syritsyn, Sergey and Yoon,
  Boram}}}, \bibinfo{journal}{PoS} \textbf{\bibinfo{volume}{LATTICE2015}},
  \bibinfo{pages}{117} (\bibinfo{year}{2016}).

\bibitem[{\citenamefont{Ji}(2013)}]{Ji:2013dva}
\bibinfo{author}{\bibfnamefont{X.}~\bibnamefont{Ji}}, \bibinfo{journal}{Phys.
  Rev. Lett.} \textbf{\bibinfo{volume}{110}}, \bibinfo{pages}{262002}
  (\bibinfo{year}{2013}), \eprint{1305.1539}.

\bibitem[{\citenamefont{Ma and Qiu}(2014)}]{Ma:2014jla}
\bibinfo{author}{\bibfnamefont{Y.-Q.} \bibnamefont{Ma}} \bibnamefont{and}
  \bibinfo{author}{\bibfnamefont{J.-W.} \bibnamefont{Qiu}}
  (\bibinfo{year}{2014}), \eprint{1404.6860}.

\bibitem[{\citenamefont{Radyushkin}(2017)}]{Radyushkin:2017cyf}
\bibinfo{author}{\bibfnamefont{A.~V.} \bibnamefont{Radyushkin}}
  (\bibinfo{year}{2017}), \eprint{1705.01488}.

\bibitem[{\citenamefont{Lin et~al.}(2015)\citenamefont{Lin, Chen, Cohen, and
  Ji}}]{Lin:2014zya}
\bibinfo{author}{\bibfnamefont{H.-W.} \bibnamefont{Lin}},
  \bibinfo{author}{\bibfnamefont{J.-W.} \bibnamefont{Chen}},
  \bibinfo{author}{\bibfnamefont{S.~D.} \bibnamefont{Cohen}}, \bibnamefont{and}
  \bibinfo{author}{\bibfnamefont{X.}~\bibnamefont{Ji}}, \bibinfo{journal}{Phys.
  Rev.} \textbf{\bibinfo{volume}{D91}}, \bibinfo{pages}{054510}
  (\bibinfo{year}{2015}), \eprint{1402.1462}.

\bibitem[{\citenamefont{Alexandrou et~al.}(2015)\citenamefont{Alexandrou,
  Cichy, Drach, Garcia-Ramos, Hadjiyiannakou, Jansen, Steffens, and
  Wiese}}]{Alexandrou:2015rja}
\bibinfo{author}{\bibfnamefont{C.}~\bibnamefont{Alexandrou}},
  \bibinfo{author}{\bibfnamefont{K.}~\bibnamefont{Cichy}},
  \bibinfo{author}{\bibfnamefont{V.}~\bibnamefont{Drach}},
  \bibinfo{author}{\bibfnamefont{E.}~\bibnamefont{Garcia-Ramos}},
  \bibinfo{author}{\bibfnamefont{K.}~\bibnamefont{Hadjiyiannakou}},
  \bibinfo{author}{\bibfnamefont{K.}~\bibnamefont{Jansen}},
  \bibinfo{author}{\bibfnamefont{F.}~\bibnamefont{Steffens}}, \bibnamefont{and}
  \bibinfo{author}{\bibfnamefont{C.}~\bibnamefont{Wiese}},
  \bibinfo{journal}{Phys. Rev.} \textbf{\bibinfo{volume}{D92}},
  \bibinfo{pages}{014502} (\bibinfo{year}{2015}), \eprint{1504.07455}.

\bibitem[{\citenamefont{Orginos et~al.}(2017)\citenamefont{Orginos, Radyushkin,
  Karpie, and Zafeiropoulos}}]{Orginos:2017kos}
\bibinfo{author}{\bibfnamefont{K.}~\bibnamefont{Orginos}},
  \bibinfo{author}{\bibfnamefont{A.}~\bibnamefont{Radyushkin}},
  \bibinfo{author}{\bibfnamefont{J.}~\bibnamefont{Karpie}}, \bibnamefont{and}
  \bibinfo{author}{\bibfnamefont{S.}~\bibnamefont{Zafeiropoulos}}
  (\bibinfo{year}{2017}), \eprint{1706.05373}.

\bibitem[{\citenamefont{Hoodbhoy et~al.}(1989)\citenamefont{Hoodbhoy, Jaffe,
  and Manohar}}]{Hoodbhoy:1988am}
\bibinfo{author}{\bibfnamefont{P.}~\bibnamefont{Hoodbhoy}},
  \bibinfo{author}{\bibfnamefont{R.~L.} \bibnamefont{Jaffe}}, \bibnamefont{and}
  \bibinfo{author}{\bibfnamefont{A.}~\bibnamefont{Manohar}},
  \bibinfo{journal}{Nucl. Phys.} \textbf{\bibinfo{volume}{B312}},
  \bibinfo{pages}{571} (\bibinfo{year}{1989}).

\bibitem[{\citenamefont{Meyer and Negele}(2008)}]{Meyer:2007tm}
\bibinfo{author}{\bibfnamefont{H.~B.} \bibnamefont{Meyer}} \bibnamefont{and}
  \bibinfo{author}{\bibfnamefont{J.~W.} \bibnamefont{Negele}},
  \bibinfo{journal}{Phys. Rev.} \textbf{\bibinfo{volume}{D77}},
  \bibinfo{pages}{037501} (\bibinfo{year}{2008}), \eprint{0707.3225}.

\bibitem[{\citenamefont{Alexandrou et~al.}(2014)\citenamefont{Alexandrou,
  Drach, Hadjiyiannakou, Jansen, Kostrzewa, and Wiese}}]{Alexandrou:2013tfa}
\bibinfo{author}{\bibfnamefont{C.}~\bibnamefont{Alexandrou}},
  \bibinfo{author}{\bibfnamefont{V.}~\bibnamefont{Drach}},
  \bibinfo{author}{\bibfnamefont{K.}~\bibnamefont{Hadjiyiannakou}},
  \bibinfo{author}{\bibfnamefont{K.}~\bibnamefont{Jansen}},
  \bibinfo{author}{\bibfnamefont{B.}~\bibnamefont{Kostrzewa}},
  \bibnamefont{and} \bibinfo{author}{\bibfnamefont{C.}~\bibnamefont{Wiese}},
  \bibinfo{journal}{PoS} \textbf{\bibinfo{volume}{LATTICE2013}},
  \bibinfo{pages}{289} (\bibinfo{year}{2014}), \eprint{1311.3174}.

\bibitem[{\citenamefont{Horsley et~al.}(2012)\citenamefont{Horsley, Millo,
  Nakamura, Perlt, Pleiter, Rakow, Schierholz, Schiller, Winter, and
  Zanotti}}]{Horsley:2012pz}
\bibinfo{author}{\bibfnamefont{R.}~\bibnamefont{Horsley}},
  \bibinfo{author}{\bibfnamefont{R.}~\bibnamefont{Millo}},
  \bibinfo{author}{\bibfnamefont{Y.}~\bibnamefont{Nakamura}},
  \bibinfo{author}{\bibfnamefont{H.}~\bibnamefont{Perlt}},
  \bibinfo{author}{\bibfnamefont{D.}~\bibnamefont{Pleiter}},
  \bibinfo{author}{\bibfnamefont{P.~E.~L.} \bibnamefont{Rakow}},
  \bibinfo{author}{\bibfnamefont{G.}~\bibnamefont{Schierholz}},
  \bibinfo{author}{\bibfnamefont{A.}~\bibnamefont{Schiller}},
  \bibinfo{author}{\bibfnamefont{F.}~\bibnamefont{Winter}}, \bibnamefont{and}
  \bibinfo{author}{\bibfnamefont{J.~M.} \bibnamefont{Zanotti}}
  (\bibinfo{collaboration}{UKQCD, QCDSF}), \bibinfo{journal}{Phys. Lett.}
  \textbf{\bibinfo{volume}{B714}}, \bibinfo{pages}{312} (\bibinfo{year}{2012}),
  \eprint{1205.6410}.

\bibitem[{\citenamefont{Alexandrou et~al.}(2017)\citenamefont{Alexandrou,
  Constantinou, Hadjiyiannakou, Jansen, Kallidonis, Koutsou, Avilés-Casco, and
  Wiese}}]{Alexandrou:2017oeh}
\bibinfo{author}{\bibfnamefont{C.}~\bibnamefont{Alexandrou}},
  \bibinfo{author}{\bibfnamefont{M.}~\bibnamefont{Constantinou}},
  \bibinfo{author}{\bibfnamefont{K.}~\bibnamefont{Hadjiyiannakou}},
  \bibinfo{author}{\bibfnamefont{K.}~\bibnamefont{Jansen}},
  \bibinfo{author}{\bibfnamefont{C.}~\bibnamefont{Kallidonis}},
  \bibinfo{author}{\bibfnamefont{G.}~\bibnamefont{Koutsou}},
  \bibinfo{author}{\bibfnamefont{A.~V.} \bibnamefont{Avilés-Casco}},
  \bibnamefont{and} \bibinfo{author}{\bibfnamefont{C.}~\bibnamefont{Wiese}}
  (\bibinfo{year}{2017}), \eprint{1706.02973}.

\bibitem[{\citenamefont{Yang et~al.}(2017)\citenamefont{Yang, Sufian,
  Alexandru, Draper, Glatzmaier, Liu, and Zhao}}]{Yang:2016plb}
\bibinfo{author}{\bibfnamefont{Y.-B.} \bibnamefont{Yang}},
  \bibinfo{author}{\bibfnamefont{R.~S.} \bibnamefont{Sufian}},
  \bibinfo{author}{\bibfnamefont{A.}~\bibnamefont{Alexandru}},
  \bibinfo{author}{\bibfnamefont{T.}~\bibnamefont{Draper}},
  \bibinfo{author}{\bibfnamefont{M.~J.} \bibnamefont{Glatzmaier}},
  \bibinfo{author}{\bibfnamefont{K.-F.} \bibnamefont{Liu}}, \bibnamefont{and}
  \bibinfo{author}{\bibfnamefont{Y.}~\bibnamefont{Zhao}},
  \bibinfo{journal}{Phys. Rev. Lett.} \textbf{\bibinfo{volume}{118}},
  \bibinfo{pages}{102001} (\bibinfo{year}{2017}), \eprint{1609.05937}.

\bibitem[{\citenamefont{Detmold and Shanahan}(2016)}]{Detmold:2016gpy}
\bibinfo{author}{\bibfnamefont{W.}~\bibnamefont{Detmold}} \bibnamefont{and}
  \bibinfo{author}{\bibfnamefont{P.~E.} \bibnamefont{Shanahan}},
  \bibinfo{journal}{Phys. Rev.} \textbf{\bibinfo{volume}{D94}},
  \bibinfo{pages}{014507} (\bibinfo{year}{2016}), \eprint{1606.04505}.

\bibitem[{\citenamefont{Detmold et~al.}(2017)\citenamefont{Detmold, Pefkou, and
  Shanahan}}]{Detmold:2017oqb}
\bibinfo{author}{\bibfnamefont{W.}~\bibnamefont{Detmold}},
  \bibinfo{author}{\bibfnamefont{D.}~\bibnamefont{Pefkou}}, \bibnamefont{and}
  \bibinfo{author}{\bibfnamefont{P.~E.} \bibnamefont{Shanahan}},
  \bibinfo{journal}{Phys. Rev.} \textbf{\bibinfo{volume}{D95}},
  \bibinfo{pages}{114515} (\bibinfo{year}{2017}), \eprint{1703.08220}.

\bibitem[{\citenamefont{Belitsky and Mueller}(2000)}]{Belitsky:2000jk}
\bibinfo{author}{\bibfnamefont{A.~V.} \bibnamefont{Belitsky}} \bibnamefont{and}
  \bibinfo{author}{\bibfnamefont{D.}~\bibnamefont{Mueller}},
  \bibinfo{journal}{Phys. Lett.} \textbf{\bibinfo{volume}{B486}},
  \bibinfo{pages}{369} (\bibinfo{year}{2000}), \eprint{hep-ph/0005028}.

\bibitem[{\citenamefont{{Orginos, Kostas and Parre\~no, Assumpta and Savage,
  Martin J. and Beane, Silas R. and Chang, Emmanuel and Detmold,
  William}}(2015)}]{Orginos:2015aya}
\bibinfo{author}{\bibnamefont{{Orginos, Kostas and Parre\~no, Assumpta and
  Savage, Martin J. and Beane, Silas R. and Chang, Emmanuel and Detmold,
  William}}}, \bibinfo{journal}{Phys. Rev.} \textbf{\bibinfo{volume}{D92}},
  \bibinfo{pages}{114512} (\bibinfo{year}{2015}), \eprint{1508.07583}.

\bibitem[{\citenamefont{{Beane, S. R. and Chang, E. and Cohen, S. D. and
  Detmold, William and Lin, H. W. and Luu, T. C. and Orginos, K. and Parre\~no,
  A. and Savage, M. J. and Walker-Loud, A.}}(2013)}]{Beane:2012vq}
\bibinfo{author}{\bibnamefont{{Beane, S. R. and Chang, E. and Cohen, S. D. and
  Detmold, William and Lin, H. W. and Luu, T. C. and Orginos, K. and Parre\~no,
  A. and Savage, M. J. and Walker-Loud, A.}}}
  (\bibinfo{collaboration}{NPLQCD}), \bibinfo{journal}{Phys. Rev.}
  \textbf{\bibinfo{volume}{D87}}, \bibinfo{pages}{034506}
  (\bibinfo{year}{2013}), \eprint{1206.5219}.

\bibitem[{\citenamefont{Taneja et~al.}(2012)\citenamefont{Taneja, Kathuria,
  Liuti, and Goldstein}}]{Taneja:2011sy}
\bibinfo{author}{\bibfnamefont{S.~K.} \bibnamefont{Taneja}},
  \bibinfo{author}{\bibfnamefont{K.}~\bibnamefont{Kathuria}},
  \bibinfo{author}{\bibfnamefont{S.}~\bibnamefont{Liuti}}, \bibnamefont{and}
  \bibinfo{author}{\bibfnamefont{G.~R.} \bibnamefont{Goldstein}},
  \bibinfo{journal}{Phys. Rev.} \textbf{\bibinfo{volume}{D86}},
  \bibinfo{pages}{036008} (\bibinfo{year}{2012}), \eprint{1101.0581}.

\bibitem[{\citenamefont{{L\"uscher, M. and Weisz, P.}}(1985)}]{Luscher:1984xn}
\bibinfo{author}{\bibnamefont{{L\"uscher, M. and Weisz, P.}}},
  \bibinfo{journal}{Commun. Math. Phys.} \textbf{\bibinfo{volume}{97}},
  \bibinfo{pages}{59} (\bibinfo{year}{1985}), \bibinfo{note}{[Erratum: Commun.
  Math. Phys.98,433(1985)]}.

\bibitem[{\citenamefont{Morningstar and Peardon}(2004)}]{Morningstar:2003gk}
\bibinfo{author}{\bibfnamefont{C.}~\bibnamefont{Morningstar}} \bibnamefont{and}
  \bibinfo{author}{\bibfnamefont{M.~J.} \bibnamefont{Peardon}},
  \bibinfo{journal}{Phys. Rev.} \textbf{\bibinfo{volume}{D69}},
  \bibinfo{pages}{054501} (\bibinfo{year}{2004}), \eprint{hep-lat/0311018}.

\bibitem[{\citenamefont{Sheikholeslami and
  Wohlert}(1985)}]{Sheikholeslami:1985ij}
\bibinfo{author}{\bibfnamefont{B.}~\bibnamefont{Sheikholeslami}}
  \bibnamefont{and} \bibinfo{author}{\bibfnamefont{R.}~\bibnamefont{Wohlert}},
  \bibinfo{journal}{Nucl. Phys.} \textbf{\bibinfo{volume}{B259}},
  \bibinfo{pages}{572} (\bibinfo{year}{1985}).

\bibitem[{\citenamefont{Detmold and Orginos}(2013)}]{Detmold:2012eu}
\bibinfo{author}{\bibfnamefont{W.}~\bibnamefont{Detmold}} \bibnamefont{and}
  \bibinfo{author}{\bibfnamefont{K.}~\bibnamefont{Orginos}},
  \bibinfo{journal}{Phys. Rev.} \textbf{\bibinfo{volume}{D87}},
  \bibinfo{pages}{114512} (\bibinfo{year}{2013}), \eprint{1207.1452}.

\bibitem[{\citenamefont{Basak et~al.}(2005)\citenamefont{Basak, Edwards,
  Fleming, Heller, Morningstar, Richards, Sato, and Wallace}}]{Basak:2005ir}
\bibinfo{author}{\bibfnamefont{S.}~\bibnamefont{Basak}},
  \bibinfo{author}{\bibfnamefont{R.}~\bibnamefont{Edwards}},
  \bibinfo{author}{\bibfnamefont{G.~T.} \bibnamefont{Fleming}},
  \bibinfo{author}{\bibfnamefont{U.~M.} \bibnamefont{Heller}},
  \bibinfo{author}{\bibfnamefont{C.}~\bibnamefont{Morningstar}},
  \bibinfo{author}{\bibfnamefont{D.}~\bibnamefont{Richards}},
  \bibinfo{author}{\bibfnamefont{I.}~\bibnamefont{Sato}}, \bibnamefont{and}
  \bibinfo{author}{\bibfnamefont{S.~J.} \bibnamefont{Wallace}}
  (\bibinfo{collaboration}{Lattice Hadron Physics (LHPC)}),
  \bibinfo{journal}{Phys. Rev.} \textbf{\bibinfo{volume}{D72}},
  \bibinfo{pages}{074501} (\bibinfo{year}{2005}), \eprint{hep-lat/0508018}.

\bibitem[{\citenamefont{Wagman et~al.}(2017)\citenamefont{Wagman, Winter,
  Chang, Davoudi, Detmold, Orginos, Savage, and Shanahan}}]{Wagman:2017tmp}
\bibinfo{author}{\bibfnamefont{M.~L.} \bibnamefont{Wagman}},
  \bibinfo{author}{\bibfnamefont{F.}~\bibnamefont{Winter}},
  \bibinfo{author}{\bibfnamefont{E.}~\bibnamefont{Chang}},
  \bibinfo{author}{\bibfnamefont{Z.}~\bibnamefont{Davoudi}},
  \bibinfo{author}{\bibfnamefont{W.}~\bibnamefont{Detmold}},
  \bibinfo{author}{\bibfnamefont{K.}~\bibnamefont{Orginos}},
  \bibinfo{author}{\bibfnamefont{M.~J.} \bibnamefont{Savage}},
  \bibnamefont{and} \bibinfo{author}{\bibfnamefont{P.~E.}
  \bibnamefont{Shanahan}} (\bibinfo{year}{2017}), \eprint{1706.06550}.

\bibitem[{\citenamefont{Beane et~al.}(2017)}]{Beane:2017edf}
\bibinfo{author}{\bibfnamefont{S.~R.} \bibnamefont{Beane}} \bibnamefont{et~al.}
  (\bibinfo{year}{2017}), \eprint{1705.09239}.

\bibitem[{\citenamefont{{L\"uscher, Martin}}(2010)}]{Luscher:2010iy}
\bibinfo{author}{\bibnamefont{{L\"uscher, Martin}}}, \bibinfo{journal}{JHEP}
  \textbf{\bibinfo{volume}{08}}, \bibinfo{pages}{071} (\bibinfo{year}{2010}),
  \bibinfo{note}{[Erratum: JHEP03,092(2014)]}, \eprint{1006.4518}.

\bibitem[{\citenamefont{Hasenfratz and Knechtli}(2001)}]{Hasenfratz:2001hp}
\bibinfo{author}{\bibfnamefont{A.}~\bibnamefont{Hasenfratz}} \bibnamefont{and}
  \bibinfo{author}{\bibfnamefont{F.}~\bibnamefont{Knechtli}},
  \bibinfo{journal}{Phys. Rev.} \textbf{\bibinfo{volume}{D64}},
  \bibinfo{pages}{034504} (\bibinfo{year}{2001}), \eprint{hep-lat/0103029}.

\bibitem[{\citenamefont{{G\"ockeler, M. and Horsley, R. and Ilgenfritz,
  Ernst-Michael and Perlt, H. and Rakow, Paul E. L. and Schierholz, G. and
  Schiller, A.}}(1996)}]{Gockeler:1996mu}
\bibinfo{author}{\bibnamefont{{G\"ockeler, M. and Horsley, R. and Ilgenfritz,
  Ernst-Michael and Perlt, H. and Rakow, Paul E. L. and Schierholz, G. and
  Schiller, A.}}}, \bibinfo{journal}{Phys. Rev.}
  \textbf{\bibinfo{volume}{D54}}, \bibinfo{pages}{5705} (\bibinfo{year}{1996}),
  \eprint{hep-lat/9602029}.

\bibitem[{\citenamefont{Alexandrou et~al.}(2016)\citenamefont{Alexandrou,
  Constantinou, Hadjiyiannakou, Jansen, Panagopoulos, and
  Wiese}}]{Alexandrou:2016ekb}
\bibinfo{author}{\bibfnamefont{C.}~\bibnamefont{Alexandrou}},
  \bibinfo{author}{\bibfnamefont{M.}~\bibnamefont{Constantinou}},
  \bibinfo{author}{\bibfnamefont{K.}~\bibnamefont{Hadjiyiannakou}},
  \bibinfo{author}{\bibfnamefont{K.}~\bibnamefont{Jansen}},
  \bibinfo{author}{\bibfnamefont{H.}~\bibnamefont{Panagopoulos}},
  \bibnamefont{and} \bibinfo{author}{\bibfnamefont{C.}~\bibnamefont{Wiese}}
  (\bibinfo{year}{2016}), \eprint{1611.06901}.

\bibitem[{\citenamefont{Dulat et~al.}(2016)\citenamefont{Dulat, Hou, Gao,
  Guzzi, Huston, Nadolsky, Pumplin, Schmidt, Stump, and Yuan}}]{Dulat:2015mca}
\bibinfo{author}{\bibfnamefont{S.}~\bibnamefont{Dulat}},
  \bibinfo{author}{\bibfnamefont{T.-J.} \bibnamefont{Hou}},
  \bibinfo{author}{\bibfnamefont{J.}~\bibnamefont{Gao}},
  \bibinfo{author}{\bibfnamefont{M.}~\bibnamefont{Guzzi}},
  \bibinfo{author}{\bibfnamefont{J.}~\bibnamefont{Huston}},
  \bibinfo{author}{\bibfnamefont{P.}~\bibnamefont{Nadolsky}},
  \bibinfo{author}{\bibfnamefont{J.}~\bibnamefont{Pumplin}},
  \bibinfo{author}{\bibfnamefont{C.}~\bibnamefont{Schmidt}},
  \bibinfo{author}{\bibfnamefont{D.}~\bibnamefont{Stump}}, \bibnamefont{and}
  \bibinfo{author}{\bibfnamefont{C.~P.} \bibnamefont{Yuan}},
  \bibinfo{journal}{Phys. Rev.} \textbf{\bibinfo{volume}{D93}},
  \bibinfo{pages}{033006} (\bibinfo{year}{2016}), \eprint{1506.07443}.

\bibitem[{\citenamefont{Eskola et~al.}(2017)\citenamefont{Eskola, Paakkinen,
  Paukkunen, and Salgado}}]{Eskola:2016oht}
\bibinfo{author}{\bibfnamefont{K.~J.} \bibnamefont{Eskola}},
  \bibinfo{author}{\bibfnamefont{P.}~\bibnamefont{Paakkinen}},
  \bibinfo{author}{\bibfnamefont{H.}~\bibnamefont{Paukkunen}},
  \bibnamefont{and} \bibinfo{author}{\bibfnamefont{C.~A.}
  \bibnamefont{Salgado}}, \bibinfo{journal}{Eur. Phys. J.}
  \textbf{\bibinfo{volume}{C77}}, \bibinfo{pages}{163} (\bibinfo{year}{2017}),
  \eprint{1612.05741}.

\bibitem[{\citenamefont{Kovarik et~al.}(2016)}]{Kovarik:2015cma}
\bibinfo{author}{\bibfnamefont{K.}~\bibnamefont{Kovarik}} \bibnamefont{et~al.},
  \bibinfo{journal}{Phys. Rev.} \textbf{\bibinfo{volume}{D93}}
  (\bibinfo{year}{2016}), \eprint{1509.00792}.

\bibitem[{\citenamefont{Kaplan and Savage}(1996)}]{Kaplan:1995yg}
\bibinfo{author}{\bibfnamefont{D.~B.} \bibnamefont{Kaplan}} \bibnamefont{and}
  \bibinfo{author}{\bibfnamefont{M.~J.} \bibnamefont{Savage}},
  \bibinfo{journal}{Phys. Lett.} \textbf{\bibinfo{volume}{B365}},
  \bibinfo{pages}{244} (\bibinfo{year}{1996}), \eprint{hep-ph/9509371}.

\bibitem[{\citenamefont{Mattis and Mukherjee}(1988)}]{Mattis:1988hg}
\bibinfo{author}{\bibfnamefont{M.~P.} \bibnamefont{Mattis}} \bibnamefont{and}
  \bibinfo{author}{\bibfnamefont{M.}~\bibnamefont{Mukherjee}},
  \bibinfo{journal}{Phys. Rev. Lett.} \textbf{\bibinfo{volume}{61}},
  \bibinfo{pages}{1344} (\bibinfo{year}{1988}).

\bibitem[{\citenamefont{Dai et~al.}(1996)\citenamefont{Dai, Dashen, Jenkins,
  and Manohar}}]{Dai:1995zg}
\bibinfo{author}{\bibfnamefont{J.}~\bibnamefont{Dai}},
  \bibinfo{author}{\bibfnamefont{R.~F.} \bibnamefont{Dashen}},
  \bibinfo{author}{\bibfnamefont{E.~E.} \bibnamefont{Jenkins}},
  \bibnamefont{and} \bibinfo{author}{\bibfnamefont{A.~V.}
  \bibnamefont{Manohar}}, \bibinfo{journal}{Phys. Rev.}
  \textbf{\bibinfo{volume}{D53}}, \bibinfo{pages}{273} (\bibinfo{year}{1996}),
  \eprint{hep-ph/9506273}.

\bibitem[{\citenamefont{Dashen et~al.}(1995)\citenamefont{Dashen, Jenkins, and
  Manohar}}]{Dashen:1994qi}
\bibinfo{author}{\bibfnamefont{R.~F.} \bibnamefont{Dashen}},
  \bibinfo{author}{\bibfnamefont{E.~E.} \bibnamefont{Jenkins}},
  \bibnamefont{and} \bibinfo{author}{\bibfnamefont{A.~V.}
  \bibnamefont{Manohar}}, \bibinfo{journal}{Phys. Rev.}
  \textbf{\bibinfo{volume}{D51}}, \bibinfo{pages}{3697} (\bibinfo{year}{1995}),
  \eprint{hep-ph/9411234}.

\bibitem[{\citenamefont{Dashen et~al.}(1994)\citenamefont{Dashen, Jenkins, and
  Manohar}}]{Dashen:1993jt}
\bibinfo{author}{\bibfnamefont{R.~F.} \bibnamefont{Dashen}},
  \bibinfo{author}{\bibfnamefont{E.~E.} \bibnamefont{Jenkins}},
  \bibnamefont{and} \bibinfo{author}{\bibfnamefont{A.~V.}
  \bibnamefont{Manohar}}, \bibinfo{journal}{Phys. Rev.}
  \textbf{\bibinfo{volume}{D49}}, \bibinfo{pages}{4713} (\bibinfo{year}{1994}),
  \bibinfo{note}{[Erratum: Phys. Rev.D51,2489(1995)]}, \eprint{hep-ph/9310379}.

\bibitem[{\citenamefont{Beane et~al.}(2013)}]{Beane:2013br}
\bibinfo{author}{\bibfnamefont{S.~R.} \bibnamefont{Beane}} \bibnamefont{et~al.}
  (\bibinfo{collaboration}{NPLQCD}), \bibinfo{journal}{Phys. Rev.}
  \textbf{\bibinfo{volume}{C88}}, \bibinfo{pages}{024003}
  (\bibinfo{year}{2013}), \eprint{1301.5790}.

\bibitem[{\citenamefont{Chen and Detmold}(2005)}]{Chen:2004zx}
\bibinfo{author}{\bibfnamefont{J.-W.} \bibnamefont{Chen}} \bibnamefont{and}
  \bibinfo{author}{\bibfnamefont{W.}~\bibnamefont{Detmold}},
  \bibinfo{journal}{Phys. Lett.} \textbf{\bibinfo{volume}{B625}},
  \bibinfo{pages}{165} (\bibinfo{year}{2005}), \eprint{hep-ph/0412119}.

\bibitem[{\citenamefont{Beane and Savage}(2005)}]{Beane:2004xf}
\bibinfo{author}{\bibfnamefont{S.~R.} \bibnamefont{Beane}} \bibnamefont{and}
  \bibinfo{author}{\bibfnamefont{M.~J.} \bibnamefont{Savage}},
  \bibinfo{journal}{Nucl. Phys.} \textbf{\bibinfo{volume}{A761}},
  \bibinfo{pages}{259} (\bibinfo{year}{2005}), \eprint{nucl-th/0412025}.

\bibitem[{\citenamefont{Chen et~al.}(2016)\citenamefont{Chen, Detmold, Lynn,
  and Schwenk}}]{Chen:2016bde}
\bibinfo{author}{\bibfnamefont{J.-W.} \bibnamefont{Chen}},
  \bibinfo{author}{\bibfnamefont{W.}~\bibnamefont{Detmold}},
  \bibinfo{author}{\bibfnamefont{J.~E.} \bibnamefont{Lynn}}, \bibnamefont{and}
  \bibinfo{author}{\bibfnamefont{A.}~\bibnamefont{Schwenk}}
  (\bibinfo{year}{2016}), \eprint{1607.03065}.

\bibitem[{\citenamefont{Edwards and Jo\'o}(2005)}]{Edwards:2004sx}
\bibinfo{author}{\bibfnamefont{R.~G.} \bibnamefont{Edwards}} \bibnamefont{and}
  \bibinfo{author}{\bibfnamefont{B.}~\bibnamefont{Jo\'o}}
  (\bibinfo{collaboration}{SciDAC Collaboration, LHPC Collaboration, UKQCD
  Collaboration}), \bibinfo{journal}{Nucl.Phys.Proc.Suppl.}
  \textbf{\bibinfo{volume}{140}}, \bibinfo{pages}{832} (\bibinfo{year}{2005}),
  \eprint{hep-lat/0409003}.

\bibitem[{\citenamefont{Clark et~al.}(2010)\citenamefont{Clark, Babich, Barros,
  Brower, and Rebbi}}]{Clark:2009wm}
\bibinfo{author}{\bibfnamefont{M.~A.} \bibnamefont{Clark}},
  \bibinfo{author}{\bibfnamefont{R.}~\bibnamefont{Babich}},
  \bibinfo{author}{\bibfnamefont{K.}~\bibnamefont{Barros}},
  \bibinfo{author}{\bibfnamefont{R.~C.} \bibnamefont{Brower}},
  \bibnamefont{and} \bibinfo{author}{\bibfnamefont{C.}~\bibnamefont{Rebbi}},
  \bibinfo{journal}{Comput. Phys. Commun.} \textbf{\bibinfo{volume}{181}},
  \bibinfo{pages}{1517} (\bibinfo{year}{2010}), \eprint{0911.3191}.

\bibitem[{\citenamefont{Babich et~al.}(2011)\citenamefont{Babich, Clark, Jo\'o,
  Shi, Brower, and Gottlieb}}]{Babich:2011np}
\bibinfo{author}{\bibfnamefont{R.}~\bibnamefont{Babich}},
  \bibinfo{author}{\bibfnamefont{M.~A.} \bibnamefont{Clark}},
  \bibinfo{author}{\bibfnamefont{B.}~\bibnamefont{Jo\'o}},
  \bibinfo{author}{\bibfnamefont{G.}~\bibnamefont{Shi}},
  \bibinfo{author}{\bibfnamefont{R.~C.} \bibnamefont{Brower}},
  \bibnamefont{and} \bibinfo{author}{\bibfnamefont{S.}~\bibnamefont{Gottlieb}},
  in \emph{\bibinfo{booktitle}{{SC11 International Conference for High
  Performance Computing, Networking, Storage and Analysis Seattle, Washington,
  November 12-18, 2011}}} (\bibinfo{year}{2011}), \eprint{1109.2935},
  \urlprefix\url{https://inspirehep.net/record/927455/files/arXiv:1109.2935.pdf}.

\end{thebibliography}

%%%%%%%%%%%%%%%%%%%%%%%%%%%%%%%%%%%%%%%%%%%%%%%%%%

\end{document}